\documentclass[11pt,letterpaper,onecolumn]{article}
\usepackage{indentfirst}
\usepackage{url}
\usepackage[top=2cm,bottom=2cm]{geometry}
\usepackage{caption}
\usepackage{subcaption}
\newcommand{\A}{\mathbb{A}} 
\newcommand{\p}{\partial}

\usepackage{answers}
\usepackage{authblk}
\usepackage{cancel}
\usepackage{setspace}
\usepackage{blindtext, rotating}
\usepackage{graphicx}
\usepackage{enumitem}
\usepackage{textcomp}
\usepackage{multicol}
\usepackage{mathrsfs}
\usepackage{amsmath,amsthm,amssymb}
\usepackage{indentfirst}
\usepackage{gensymb}
\usepackage{mdframed}
\usepackage{tikz-cd}
\usepackage{fancyhdr}
\usepackage{tensor}
\usepackage{braket}
\usepackage{float}
\usepackage[section]{placeins}
\usepackage{subcaption} 
\usepackage[colorlinks=true, linkcolor=blue, citecolor=green]{hyperref} 

\title{\bf On stimulated scattering including harmonic generation in an unmagnetized pair plasma for arbitrary pump strengths
}

\author[1]{Vijay Patel}
\author[2]{Thomas Grismayer}
\author[1]{Samuel Sebastian}
\author[3,4]{Xinlu Xu}
\author[5]{John Palastro}
\author[1]{Paulo Alves}
\author[1,6]{Warren B. Mori}
\affil[1]{Department of Physics and Astronomy, University of California, Los Angeles, California 90095, USA}
\affil[2]{GoLP/Instituto de Plasmas e Fusao Nuclear, Instituto Superior Tecnico, Universidade de Lisboa, Lisboa 1049-001, Portugal}
\affil[3]{State Key Laboratory of Nuclear Physics and Technology, and Key Laboratory of HEDP of the Ministry of Education, CAPT, School of Physics, Peking University, Beijing 100871, China}
\affil[4]{Beijing Laser Acceleration Innovation Center, huairou, Beijing 100871, China}
\affil[5]{Laboratory for Laser Energetics (LLE), University of Rochester, 250 East River Rd., Rochester, NY 14623, USA}
\affil[6]{Department of Electrical and Computer Engineering, University of California, Los Angeles, California 90095, USA}

\begin{document}
\maketitle

\begin{abstract}
A fully nonlinear analysis for the growth rate of the stimulated backscatter instability of an intense electromagnetic (EM)  wave with arbitrary polarization in an unmagnetized pair plasma is presented. The connection of this instability to the FEL process is discussed and utilized. An intense EM wave causes an initially stationary pair plasma to drift forward with a speed near the speed of light. The resulting distribution function for both cold and warm pair plasmas is presented.  It is shown that the frequency of the reflected light is thus Doppler downshifted from that of the incident wave. For a linearly polarized EM wave there is also an additional contribution to the forward drift at its second harmonic which leads to backscattered light at all odd harmonics of the Doppler downshifted frequency for sufficiently cold pair plasmas. Fully nonlinear particle-in-cell simulations confirm the theoretical growth rates and show how the instability depends on the pair plasma temperature. The effect of the intensity profile of the incoming EM wave as well as implications for astrophysical pair plasmas is discussed.   
\end{abstract}



\section{Introduction}






    
There is much recent interest in the physics of laser plasma interactions (LPI) in pair plasmas due to how it might effect the propagation of fast radio bursts (FRB) in the magnetosphere of magnetars and thus alter terrestial observations \cite{Lyutikov1998,Sobacchi2022,Sobacchi2023, Sobacchi2024AA, Sobacchi2024PRR,lyutikov2026,Lyutikov2020}. Recently, it has been suggested that intense EM waves can be reflected in a pair plasma \cite{Sironi, Tangtartharakul} from a process that is related to the well known stimulated Brillouin scattering (SBS) instability in electron-ion plasmas \cite{forslund1975theory}. Based on the results from this article, it can also be argued that it is also the limit of SRS in a pair plasma.
We thus refer to it as simply induced backscatter rather than SBS or SRS.  Several authors have provided theoretical predictions for the growth rates of backscatter 
in a pair plasma 
for small amplitude pump strengths,  $
eA_0/mc^2\equiv a_0 < 1$  where $A_0$, 
$e$, $m$, and $c$ are the vector potential of the pump, electron charge, electron mass, and speed of light respectively. The quantity, $a_0=.85\times10^{-5}\sqrt{I(W/cm^2)}\lambda(cm)$, is referred to as the normalized vector potential where $I$ and $\lambda$ are the radiation intensity and wavelength respectively. Very recently, several groups have presented theoretical results for backscatter from arbitrary pump strengths. Supporting particle-in-cell simulations were also described in some cases\cite{Sironi,Iwamoto2026}. 

It has also been recently recognized that the growth rate of induced backscatter in a pair plasma is scales as the Pierce parameter \cite{lyutikov2026} that arises in the free electron laser instability (FEL)\cite{pellegrini2016physics}. The precise reason for this was not discussed. The physical reason for this connection was alluded to in Ref. \cite{XuMori2026} where it was shown that 
the gain of stimulated Raman scattering (SRS) in the strongly coupled regime for a short laser pulse and the FEL instability for XFEL facilities are identical when viewed from their respective Lorentz frames
. 

In SRS, a laser (EM pump) is scattered into a backscattered EM wave and a forward propagating electron plasma wave (EPW). In  the FEL instability, a relativistic electron beam propagates through an undulator, which is typically a static magnetic field structure oriented perpendicular to the propagation direction of the electron beam, and with a well-defined wavenumber in the propagation direction. If one views the FEL process in the frame of the electron beam, the undulator transforms to a quasi-EM wave that propagates towards the stationary beam electrons. In the strongly coupled regime of SRS (or SBS) the ponderomotive force of the laser dominates the bunching of the plasma and thus space charge forces are negligible. 

With this insight, it can be seen that the differences between backscatter in a pair plasma, SRS of a short-pulse laser in a strongly coupled regime, and the XFEL instability lie in what Lorentz frame one views the process and in the regime of the instability. Specifically, in Ref. \cite{XuMori2026} it was shown that the exponentiation of SRS of a short pulse laser in the strongly coupled limit is identical to the Pierce parameter times the appropriate length of the undulator. 

In this paper, we use this insight to develop the theory of ``laser'' plasma interactions in pair plasmas for arbitrary $a_0$ using the formalism from previous work 
in the free electron laser (FEL) literature \cite{Sprangle-FEL} and previous work on SRS in an electron-ion plasma\cite{decker1996evolution}. 
Specifically,  in Ref. \cite{Sprangle-FEL} SRS backscatter of a linearly polarized (LP) laser from a stationary plasma or counterpropagating relativistic electron beam was analyzed. 
They studied the strongly coupled regimes of the instability where the space charge forces in the EPW or the modulated electron beam are negligible compared to the ponderomotive force from the forward and backward scattered EM waves. However, for a pair plasma the ponderomotive force is identical on electrons and positrons so there are no space charge forces 
when the pair plasma is modulated. Thus, the physics of backscatter in a pair plasma in the presence of  electromagnetic radiation is identical to the strongly coupled limit for electron beams propagating toward an EM wave (undulator) when viewed in the beam frame, with the only difference being a factor of $\sqrt{2}$ in the plasma frequency \cite{XuMori2026, Sprangle-FEL}. 

In this article,  we connect the formalism of Esarey and Sprangle for backscatter of an electron beam drifting backwards to backscatter in a frame of an initially stationary 
pair plasma. The stationary pair plasma drifts forward in response to the radiation pressure of the incoming EM wave. The results are confirmed  in particle-in-cell simulations using OSIRIS. We also extend the treatment of Ref. \cite{Sprangle-FEL} to include circularly polarized (CP) and elliptically polarized (EP)  electromagnetic waves. The possible implications for the results to a pair plasma surrounding a magnetar will also be discussed. 


\section{Derivation of the growth rate}
As discussed above, to analyze the growth rate of stimulated backscatter of an intense EM wave impinging on an unmagnetized plasma, we utilize the formalism of the work of Esarey and Sprangle \cite{Sprangle-FEL} who considered backscatter of an arbitrarily large LP EM wave with either a stationary plasma (with immobile ions) or a drifting electron beam with $\gamma_0$. Here, we are interested in backscatter from a pair plasma that is initially stationary. We note that their results can be extended to an initially stationary e-beam by setting $\gamma_0=1$, but in order to extend the analysis to general polarizations, we start by assuming the pair plasma is at rest and provide details of the derivation in this article. 

Their formalism requires assuming that the forward and backward propagating EM waves move at the speed of light and finding an appropriate nonlinear equilibrium to perturb about. When an EM wave  impinges on a pair plasma with an arbitrary initial drift, the motion of the electrons and positrons are known analytically for plane waves  (one-dimensional limit) moving at the speed of light (valid for a sufficiently low density plasma). 
General expressions for an equilibrium for EM waves with phase velocities that depend on the plasma density can also be obtained but this is challenging for linearly polarized EM waves. We consider this for a circularly polarized EM wave in the Appendix.

When the pair plasma is initially at rest, in the presence of the EM wave, both the electrons and positrons move forward, as well as oscillate transversely. For a circularly polarized wave the pair plasma moves forward with a constant drift, while for a linearly polarized wave it moves forward with an average drift, as well as with an oscillation at twice the frequency of the laser with respect to the variable $\xi\equiv(z-ct)$. As we show below, this leads to reflected light at a Doppler downshifted frequency due the forward drift as well as at odd harmonics of the Doppler downshifted frequency for linearly polarized light due to the additional axial motion at the second harmonic.

To determine if the forward going pump and reflected light feed back on each other to drive an instability, we use the equilibrium of an incident EM wave and the associated self-consistent motion of a  pair plasma that started at rest as the nonlinear equilibrium. 
A discussion for how to include finite temperature effects is given at the end of this section and finite temperature effects are studied using the PIC simulations.\\
\indent The fully relativistic two-fluid and maxwell equations for the pair plasma in the presence of a linearly or circularly polarized pump wave propagating in the $\hat z$ direction are given as follows:
\begin{align}
    \p_t(\gamma_\alpha n_\alpha) + \p_z(n_\alpha u_{\alpha,z}) &= 0\label{eq:1}\\
    (\p_t+ \frac{u_{\alpha,z}}{\gamma_\alpha}\p_z) \mathbf{u}_{\alpha,\perp} &= \frac{-q_\alpha}{m_\alpha c}(\p_t + \frac{u_{\alpha,z}}{\gamma_\alpha}\p_z)\mathbf{A_\perp}\label{eq:2}\\
    (\p_t+ \frac{u_{\alpha,z}}{\gamma_\alpha}\p_z) u_{\alpha,z} &= \frac{q_\alpha }{\gamma_\alpha m_\alpha c}(\mathbf{u}_{\alpha,\perp}\cdot c\partial_z )\mathbf{A_\perp}\label{eq:3}\\
    (\partial_t + \frac{u_{\alpha,z}}{\gamma_\alpha}\p_z)\gamma_\alpha &= -\frac{q_\alpha}{\gamma_\alpha m_\alpha c}(\mathbf{u}_{\alpha,\perp} \cdot \p_t) \mathbf{A_\perp}\label{eq:4}\\
    (\frac{1}{c^2}\p_t^2 
    -\p_x^2)\mathbf{A_\perp} &= \frac{4\pi }{c^2} \sum_\alpha q_\alpha n_\alpha \mathbf{u_{\alpha,\perp}}\label{eq:5}
\end{align}
where $\perp$ refers to directions orthogonal to the propagation direction of the EM waves, $\alpha = \pm$ denotes either an electron or positron fluid variable with $q_\alpha = \pm e$, and $m_\alpha = m_e$, while $n_\alpha$, $\mathbf{u_{\alpha,\perp}}$, $u_{\alpha,z}$, and $\mathbf{A_\perp}$ denote the proper densities, proper velocities, and vector potential respectively. Electrostatic effects have been neglected since as discussed above there is no mechanism for the radiation pressure  to generate 
charge separation in an unmagnetized pair plasma 
. Note that this is not true in the magnetized case \cite{Nishiura26,Kamijima26}. 

Several important relationships can be obtained by taking the linear combination of these equations. We define $D_{\alpha,t}\equiv \p_t+\frac{u_{\alpha,z}} {\gamma_\alpha}\p_z$, which is the total time derivative following a particle, and then pull the right hand side of Eqn. (\ref{eq:2}) to the left side and  take the sum and difference of Eqns. (\ref{eq:3}) and (\ref{eq:4}) to obtain,

\begin{align}
    D_{\alpha,t}(\frac{\mathbf{u_{\alpha,\perp}}}{c}+\frac{q_\alpha}{mc^2}\mathbf {A_\perp})=0\label{eq:tranv-mom}\\
    D_{\alpha,t}(\gamma_\alpha-\frac{u_{\alpha,z}}{c})+\frac{q_\alpha}{\gamma_\alpha m_\alpha c^2}[\frac{\mathbf u_{\alpha,\perp}}{c}\cdot(\p_t+c\p_z)]\mathbf{A_\perp}=0 \label{eq:g}\\
    D_{\alpha,t}(\gamma_\alpha+\frac{u_{\alpha,z}}{c})+\frac{q_\alpha}{\gamma_\alpha m_\alpha c^2}[\frac{\mathbf u_{\alpha,\perp}}{c}\cdot(\p_t-c\p_z)]\mathbf{A_\perp}=0 \label{eq:h}\\
\end{align}

Equation (\ref{eq:tranv-mom}) gives the conservation of transverse canonical momentum which is rigorously true for the one-dimensional scenario under consideration. On the other hand, eqns. (\ref{eq:g}) and (\ref{eq:h}) show that $\gamma-\frac{u_{\alpha,z}}{c}$ or $\gamma+\frac{u_{\alpha,z}}{c}$ are other useful quantities, as they are also conserved when the fields are only functions of $z-ct$,  in which case $\p_t+c\p_z=0$ is conserved, or of $z+ct$, in which case  $\p_t-c\p_z=0$ is conserved. 
From the conservation of transverse canonical momentum it is clearly seen that  $\mathbf{u_\perp}$ for the electrons and positrons are $\pi$ out of phase and are both proportional to $\mathbf{A_\perp}$. Thus, $\gamma-\frac{u_z}{c}$ and $\gamma+\frac{u_z}{c}$ are equal for electrons and positrons.  Consider an incident EM wave moving in the $+ \hat z$ direction which is either linearly polarized (LP), $\mathbf{A_\perp}=\hat x \A(\xi)cos[k_0\xi]$ or circularly polarized (CP) $\mathbf{A_\perp}=\hat x \A (\xi)cos[k_0\xi]\pm\hat y \A(\xi)sin[k_0\xi]$  where  $k_0\equiv\frac{\omega_0}{c}$ and - (+)  corresponds to right (left) handed CP polarized respectively.  Let the electron or positron start at rest in front of the incoming EM wave (it is straightforward to modify these expressions for particles that are not at rest in front of the EM wave such as for a moving electron beam). For these initial conditions, the conserved quantities can be combined to give

\begin{align}
    \frac{\mathbf{u_{\alpha,\perp}}}{c}=-\hat x\frac{q_\alpha}{mc^2} \A(\xi)cos[k_0\xi]\label{eq:equil1}\\
    \frac{u_{\alpha,z}}{c}= \frac{[\frac{q_\alpha\A(\xi)}{mc^2} ]^2cos^2[k_0\xi]}{2}\\
    \gamma_{\alpha}=1+\frac{[\frac{q_\alpha\A(\xi)}{mc^2} ]^2cos^2[k_0\xi]}{2}
     \label{eq:equil3}\\
\end{align}
for a LP EM wave and 
\begin{align}
    \frac{\mathbf{u_{\alpha,\perp}}}{c}=-\hat x \A (\xi)cos[k_0\xi]\mp\hat y \A(\xi)sin[k_0\xi]\\
    \frac{u_{\alpha,z}}{c}= \frac{[\frac{q_\alpha\A(\xi)}{mc^2} ]^2}{2}\\
\gamma_{\alpha}=1+\frac{[\frac{q_\alpha\A(\xi)}{mc^2} ]^2}{2}
     \\
\end{align}
for a CP EM wave. If the laser power is kept fixed then $\frac{q_\alpha\A(\xi)}{mc^2}$ is $2^{-1/2}$ smaller for the CP case such that the average $u_{\alpha,z}$ is the same for both LP and CP. We take this difference in amplitude for fixed power into account when comparing growth rates below. 

For either a LP or CP EM wave $u_z$ and $\gamma$ are the same for electrons and positrons and both quantities oscillate at the second harmonic for a LP wave. Thus, an initially stationary pair plasma will oscillate transversely and drift forward when an EM wave is incident upon it and for a LP wave the forward drift ($v_z$)  will also include oscillations at even harmonics.  The consequence of the conserved quantities will also be shown to be useful for defining a nonlinear equilibrium to perturb about. 

Recognizing that because electrons and positrons move together along $\hat z$, it can be seen that  there is no charge separation, and thus  single fluid quantities for the density and $\gamma\mp \frac{u_z}{c}$ can be utilized. Henceforth, we
use the normalized quantities, $\frac{e\mathbf{A_\perp}}{mc^2}\equiv \mathbf{a}$, $\frac{u_{\alpha,z}}{c}\equiv u_z$, $\frac{\mathbf{u_{\alpha,\perp}}}{c}\equiv \mathbf{u_\perp}$, and $\frac{n_\alpha}{n_0}$ 
where 
$n_0$ which is the initial unperturbed density for  the electrons and for the positrons.
In what follows,
we also utilize the relationships  
$u_{+,z} = -u_{-,z}\equiv u_z$, $\frac{n_+}{n_0}=\frac{n_+}{n_0}\equiv n$, and $\gamma_+=\gamma_-\equiv \gamma$ ,
to reduce Eqns. (\ref{eq:1})-(\ref{eq:5}) to:
\begin{align}
    \p_t(\gamma n) + \p_z(n u_{z}) &= 0 \label{eq:6}\\
    (\gamma\p_t+ u_z\p_z) u_{z} &= -\frac
    {1}{2}\partial_z |\mathbf a|^2\label{eq:7}\\
    (\gamma\partial_t + u_z
    \p_z)\gamma &= \frac{1}{2}\p_t |\mathbf a|^2\label{eq:8}\\
    (
    \p_t^2 
    -\p_z^2)\mathbf{a} &= -2
    \omega_p^2
    \frac{n}{n_0} \mathbf{a}\label{eq:9}
\end{align}
where 
$\omega_p^2 = 4\pi e^2 n_0/m_e$ is the plasma frequency and $n_0$ is the unperturbed proper density of the electrons or positrons. We now 
define the variables $\xi,\eta = (z\mp  t)$ 
and use the relationships $\p_t=(\p_\eta-\p_\xi)$ and $\p_z=(\p_\eta+\p_\xi)$ to transform
Eqns. (\ref{eq:6})-(\ref{eq:9}) 
to:
\begin{align}
    -\p_\xi[(\gamma - u_z) n] + \p_\eta[(\gamma+ u_z) n]  &= 0\label{eq:fl1}\\
    [(\gamma + u_z)\p_\eta - (\gamma - u_z)\p_\xi]\gamma &= - \frac{1}{2}(\p_\xi -\p_\eta) |\mathbf{a}|\label{eq:fl2}\\
    [(\gamma + u_z)\p_\eta - (\gamma - u_z)\p_\xi]u_z &= -\frac{1}{2}(\p_\xi +\p_\eta) |\mathbf{a}|^2\label{eq:fl3}\\
    (-4\p_\eta \p_\xi)\mathbf{a}&= -\epsilon 2\omega_p^2\frac{n}{n_0}\mathbf{a}.
\end{align}
where 
$\epsilon$ is a smallness parameter that is $O(\frac{\omega_p}{\omega_0})^2$. 
Following \cite{Sprangle-FEL}, we define $g,h \equiv \gamma \mp u_z$ and rewrite Eqns. (\ref{eq:fl1}) - (\ref{eq:fl3})  in terms of these new fluid variables, 
\begin{align}
    -\p_\xi(gn) + \p_\eta(hn)  &= 0\label{eq:4a}\\
    h\p_\eta h  - g\p_\xi h &= - \p_\xi |\mathbf{a}|^2 \label{eq:4b}\\
    h\p_\eta g  - g\p_\xi g  &= \p_\eta |\mathbf{a}|^2\label{eq:4c}\\
    -2\p_\eta \p_\xi\mathbf{a} &= -\frac{\omega_p^2}{c^2}\frac{n}{n_0}\mathbf{a}\label{eq:4d}
\end{align}
From here on, we implicitly assume that $\omega_p \ll \omega_0$ so that all EM waves relevant to the problem travel at the speed of light. 
We follow the notation in \cite{Sprangle-FEL} where perturbation theory for the above equations proceeds by expanding each fluid variable $Q$ as $Q(\xi, \eta) = Q^{(0)}(\xi) + Q^{(1)}(\xi, \eta)$ and $Q^{(1)} \ll Q^{(0)}$. The zeroth order stationary solutions, are obtained by setting $\p_\eta=0$ to obtain,
\begin{align}
    \p_\xi(g^{(0)}n^{(0)}) &= 0\\
    g^{(0)}\p_\xi g^{(0)} &= 0\\
    g \p_\xi h &= 
     \p_\xi {|\mathbf{a}^{(0)}|}^2
\end{align}
from which it is easily found that $g^{(0)}=constant=C_0=1$ for a pair plasma starting at rest, $n^{(0)}=constant=n_0$ (the proper density is an invariant),  and $h^{(0)} = 1+|\mathbf{a}_0|^2$. This gives $u_z^{(0)} = \frac{1}{2}|\mathbf{a}_0|^2$ and $\gamma^{(0)} = 1+\frac{1}{2}|\mathbf{a}_0|^2$. Furthermore,  $h$ can be eliminated from 
eqns. (\ref{eq:4a}) and (\ref{eq:4b}) by making use of the relations $gh = 1+|\mathbf{a}|^2$ and $\p_\xi g = -\p_\eta h$. Note that these zeroth order solutions are the same as obtained earlier from single particle dynamics. Utilizing 
these relationships the following three coupled equations are obtained
\begin{align}
    g^2\p_\xi n -(1+|\mathbf{a}|^2)\p_\eta n &= - n\p_\xi g^2\\
    g^2\p_\xi g - (1+|\mathbf{a}|^2)\p_\eta g &= - g \p_\eta a^2\\
    \p_\eta \p_\xi \mathbf{a} &=4\frac{\omega_p^2}{c^2}\frac{n}{n_0}\mathbf{a}
\end{align}
The perturbed solutions are assumed to be of the form $Q^{(1)}(\eta,\xi) = Q_1(\xi)e^{ik \eta}$. Perturbing the above equations around the equilibrium solutions gives:
\begin{align}
    (\p_\xi - ik(1 + \mathbf{a}^{(0)}(\xi)^2))g_1 &= -2 ik(\mathbf{a}^{(0)}\cdot \mathbf{a}_1)\\
    (\p_\xi - ik(1 + \mathbf{a}^{(0)}(\xi)^2))\frac{n_1}{n_0} &= -2 \p_\xi g_1\\
    (ik\p_\xi -\frac{\omega_p^2}{4c^2})\mathbf{a}_1 &= \frac{\omega_p^2}{4c^2}\mathbf{a}^{(0)} \frac{n_1}{n_0}\label{eq:a1}
\end{align}
The solutions to these equations 
are given by:
\begin{align}
    g_1 &= -2ike^{i\theta(\xi)}\int d\xi'e^{-i\theta(\xi')}(\mathbf{a}_{0}(\xi')\cdot \mathbf{a}_1(\xi'))\label{eq:h1}\\
    \frac{n_1}{n_0}&= -2 g_1(\xi) -2ik e^{i\theta(\xi)}\int d\xi' e^{-i\theta(\xi')}\mu(\xi')g_1(\xi')\label{eq:n1}
\end{align}
where $\mu(\xi) = 1+a^{(0)}(\xi)^2$ and $\theta(\xi) = k\int d\xi' \mu(\xi')$. We assume that the pump is elliptically polarized as $\mathbf{a}^{(0)}(\xi) = a_0( p_1\cos(k_0\xi)\hat{y} + p_2\sin(k_0\xi)\hat{x})=Re[a_1 (p_1\hat y- ip_2\hat x)
e^{i\Delta k \xi}]$
and the backscattered light wave amplitude has a similar form $\mathbf {a_1}(\xi)=  Re[a_1 (p_1\hat y- ip_2\hat x)
e^{i\Delta k \xi}]$
where $p_1^2+p_2^2 = 1$. The fact that the backscattered wave has a different sense of rotation in space (but not time) follows from the analysis that follows. 
Using these definitions leads to:
\begin{align}
    \mu(\xi) &= 1 + \frac{a_0^2}{2} + a_0^2\frac{\alpha}{2}\cos(2k_0\xi)\\
    \theta(\xi) &= \bar{k}\xi + a_0^2\frac{k\alpha}{4k_0}\sin(2k_0\xi)
\end{align}
where $\alpha\equiv p_1^2- p_2^2$ 
Using these relationships in Eqn. (\ref{eq:h1}) and  defining $\bar{k} \equiv (1+a_0^2/2)k$, 
leads to :
\begin{align}
    g_1(\xi) &= -2ika_0a_1e^{i\bar{k}\xi}\sum_{m,l} J_m(\alpha b)J_l(\alpha b)(-1)^le^{2imk_0\xi}\int d\xi'\bigg[\frac{p_1^2}{2}\bigg(e^{i((2l + 1) k_0 - \bar{k})\xi'} + e^{i((2l - 1) k_0 - \bar{k})\xi'}\bigg)\nonumber\\
    &-\frac{p_2^2}{2}\bigg(e^{i((2l + 1) k_0 - \bar{k})\xi'} - e^{i((2l - 1) k_0 - \bar{k})\xi'}\bigg)\bigg]e^{i\Delta k\xi'}\nonumber\\
    &= ka_0a_1\sum_{m,l}\frac{ J_m(\alpha b)(-1)^l( J_l(\alpha b)- \alpha J_{l +1}(\alpha b))}{\bar{k}-\Delta k - (2l + 1)k_0 }e^{i(\Delta k+(2(l+m) + 1) k_0 )\xi}
\end{align}
and $b \equiv \frac{ka_0^2}{4k_0}$
. The Bessel expansion $e^{ia\sin\theta} = \sum_m  J_m(a) e^{im\theta}$ has been used in the first line, and in the second line, the indices were labeled $l\mapsto -l,-l+1$ for the second and first terms in the cosine. Furthermore, $n_1$ can be evaluated similarly by inserting Eqn. (\ref{eq:h1}) into Eqn. (\ref{eq:n1}):
\begin{equation}
\begin{split}
    \frac{n_1}{n_0}&= -2g_1-4k^2a_0a_1e^{i\bar{k}\xi}\sum_{m,l} J_m(\alpha b)J_l(\alpha b)e^{2imk_0\xi}\int d\xi'\mu(\xi')\int d\xi''\exp{(i(-2lk_0-\bar{k} + \Delta k)\xi'')}\times\\
    &( p_1^2\cos(k_0\xi'')+ ip_2^2\sin(k_0\xi''))\\
    &=-2g_1(\xi)-4k^2a_0a_1e^{i\bar{k}\xi}\sum_{m,l} J_m(\alpha b_l)(J_l(\alpha b)-  \alpha J_{l + 1}(\alpha b))(-1)^le^{2imk_0\xi}\times\\
    &\int d\xi'\mu(\xi')\frac{\exp{(i((2l+1)k_0-\bar{k} + \Delta k)\xi')}}{2i((2l + 1)k_0  - \bar{k} + \Delta k)}\\
    &= -2g_1(\xi) + 2k\bar{k}a_0a_1\sum_{m,l} J_m(\alpha b)( J_l(\alpha b)-\alpha J_{l + 1}(\alpha b))(-1)^l\frac{\exp{(i((2(l + m)+1)k_0 + \Delta k)\xi)}}{((2l + 1)k_0  - \bar{k} + \Delta k)^2}\\
    &+\frac{k^2\alpha a_0^2}{4k_0}a_0a_1\sum_{m,l} J_m(\alpha b)(-1)^l\exp((i(\Delta k + 2(l+m)k_0)\xi)\times\\
    &\bigg(\frac{ \alpha J_l(\alpha b)- J_{l - 1}(\alpha b)}{(\Delta k - (2l - 1) k_0-\bar{k})(\Delta k - (2l + 1) k_0-\bar{k})}
    +\frac{ \alpha J_{l + 2}(\alpha b) -  J_{l + 1}(\alpha b)}{(\Delta k - (2l +3) k_0-\bar{k})(\Delta k - (2l + 1) k_0-\bar{k})}\bigg)
\end{split}
\end{equation}

The second term in the parenthesis in the final line contains only  
contributes 
nonresonant terms to the final expression for $n_1$, so it can be safely dropped.
The initial $g_1$ term can also be dropped because the denominator, which is small near resonances, is to the first power, not the second. 
The dispersion relation can be found by inserting the above expressions for $g_1,n_1$ into Eqn. (\ref{eq:a1}):
\begin{equation}
\begin{split}
    &(-k\Delta k - \frac{\omega_p^2}{4c^2})a_1 = \frac{\omega_p^2k\bar{k}a_0a_1}{2c^2}\sum_{m,l} J_m(\alpha b)(  J_l(\alpha b)- \alpha J_{l + 1}(\alpha b))\times \\
    &(-1)^l\frac{\exp{(i((2(l + m)+1)k_0 )\xi)}}{((2l + 1)k_0  - \bar{k} + \Delta k)^2}(p_1^2\cos(k_0\xi)
    -ip_2^2\sin(k_0\xi))\\
    \end{split}
\end{equation}
    where the right hand side  reduces to,
    \begin{equation}
\begin{split}
    &\frac{\omega_p^2k\bar{k}a_0a_1}{4c^2}\sum_{m,l}(  J_m(\alpha b)+ \alpha J_{m-1}(\alpha b))(   J_l(\alpha b)-  \alpha J_{l + 1}(\alpha b))(-1)^{l}\frac{\exp{(i(2(l + m)k_0 )\xi)}}{((2l + 1)k_0  - \bar{k} + \Delta k)}.\\
\end{split}
\end{equation}
On physical grounds, we can assume the dominant contributions come from the resonant terms where $m=-l$. Under this assumption and the use of the Bessel identity $J_{-l}(x) = (-1)^lJ_l(x)$, this reduces to, 
\begin{equation}
\begin{split}
    \sim\frac{\omega_p^2k\bar{k}a_0a_1}{4c^2}\sum_{l}\frac{(  J_l(\alpha b)-  \alpha J_{l + 1}(\alpha b))^2}{((2l + 1)k_0  - \bar{k} + \Delta k)^2}.
\end{split}
\end{equation}
This leads to 
a final form for the dispersion relation:
\begin{align}
    4k\Delta k + \frac{\omega_p^2}{c^2}
    &=-\frac{\omega_p^2}{c^2}k\bar{k}a_0^2\sum_{l}\frac{(  J_l(\alpha b)-   \alpha J_{l + 1}(\alpha b))^2}{((2l + 1)k_0  - \bar{k} + \Delta k)^2}.
\end{align}
We define $\omega/c = k + \Delta k$, and if we assume $|\Delta k| \ll k$, then exponential growth occurs near resonances of the denominator where is achieved at $\omega = (2l + 1)M_0 \omega_0$ and $M_0 = 1/(a_0^2/2 + 1)\equiv 1/\gamma_0$. 

The dispersion relation for the $l$th mode is therefore:
\begin{equation}
    \Delta k^3 =\frac{\omega_p^2 k_0M_0 b_l(  J_l(\alpha b_l)-   \alpha J_{l + 1}(\alpha b_l))^2}{c^2(1+M_0)^2}
\end{equation}
where $b_l = \frac{ka_0^2}{4k_0}= \frac{(2l + 1)M_0a_0^2}{4}=\frac{NM_0a_0^2}{4}$ where $N\equiv 2l+1$ is the harmonic number. We have adopted the notation  in \cite{Sprangle-FEL} of using  $M_0$ so that it will be easier to compare results. We note that as shown in Ref. \cite{Sprangle_harmonicsSRS_PRL} a general expression $M_0$ is $M_0=C_0^2/(1+a_0^2/2)$ where recall that $C_0=g^{(0)}=\gamma_0-p_{z0}$.

In addition, the argument of the Bessel functions are $\alpha b_l$
Therefore, for circularly polarized pumps, where $\alpha = 0$
, it follows that $J_0(\alpha b)=1$ and $ J_l(\alpha b)= 0$ for $l \neq 0$ such that only the fundamental mode is able to be resonant. For linearly polarized pumps, $\alpha = 1$
and all harmonics are able to become resonant and contribute to the instability. 

Since $a^{(1)}(\xi,\eta) \sim \exp i[(\omega/c)\eta - 2i \Delta k t]$, it follows that the growth rate $\Gamma$ is $2\Im\Delta k$ so that
\begin{equation}
    \frac{\Gamma}{k_0} =\sqrt{3}\bigg(\frac{\omega_p^2 M_0
    b_l(   J_l(\alpha b_l)-   \alpha J_{l + 1}(\alpha b_l))^2}{\omega_0^2
    (1+M_0)^2}\bigg)^{1/3}\label{eq:growth-rate}
\end{equation}
where recall that $\alpha = 1$ for LP  and $\alpha =0$ for CP from 
which it follows,
\begin{equation}
    \frac{\Gamma_{LP}}{k_0} =\sqrt{3}\bigg(\frac{\omega_p^2 M_0
    b_l(  J_l(\alpha b_l)-   \alpha J_{l + 1}(\alpha b_l))^2}{\omega_0^2
    (1+M_0)^2}\bigg)^{1/3}\label{eq:growth-rate-LP}
\end{equation}
 for each $l$ value, and

 \begin{equation}
    \frac{\Gamma_{CP}}{k_0} =\sqrt{3}\bigg(\frac{\omega_p^2 M_0
    b
    }{\omega_0^2
    (1+M_0)^2}\bigg)^{1/3}\label{eq:growth-rate-CP}
\end{equation}
At small $a_0$, $M_0 \approx 1$ and the growth rate of the fundamental for LP (and the growth rate for CP) behaves as 
\begin{equation}
    \frac{\Gamma}{k_0} \approx \sqrt{3}\bigg(\frac{\omega_p^2 }{16\omega_0^2}a_0^2\bigg)^{1/3}\label{eq:lowa0growthrate}
\end{equation}
and this is the growth rate that has been derived numerous times in the astrophysics literature {\it at small $a_0$} \cite{lyutikov2026,Ghosh2022} and is the expression for the growth of SRS of a short-laser pulse in the strongly coupled regime and XFEL in the frame of e-beam \cite{XuMori2026}.

However, for large $a_0$, $M_0 \approx 2a_0^{-2}$, therefore if the weak dependence on 
$a_0$ in the Bessel function terms is neglected, then $\Gamma$ scales as $a_0^{-2/3}$. For CP where the Bessel functions are absent the growth rate is maximum for $a_0=2$ while for LP the growth rate is 
maximum  at $a_{0} \approx 1.662$.  Thus, our expression predicts that the growth rate {\it decreases} when the motion of the pair plasma  becomes relativistic, i.e., when $a_0>1$. To understand the physical reasons behind the saturation of the growth rate
it is useful to compare the growth rate to the formula for the growth rate in Ref. \cite{Sprangle-FEL} for an ion-electron plasma that does not drift due to the massive ions, 
\begin{equation}
    \frac{\Gamma_{LP}}{k_0} =\sqrt{3}\bigg(\frac{\omega_p^2 M_0
    b_l(  J_l(\alpha b_l)- J_{l + 1}(\alpha b_l))^2}{\omega_0^2
    (1+a_0^2/2)^{1/2}}\bigg)^{1/3}\\
    =\sqrt{3}\bigg(\frac{\omega_p^2 M_0
    Na_0^2(  J_l(\alpha b_l)- J_{l + 1}(\alpha b_l))^2}{4\omega_0^2
    (1+a_0^2/2)^{3/2}}\bigg)^{1/3}.
    \label{eq:growth-rate-e-ion-LP}
\end{equation}

where $(1+a_0^2/2)^{1/2}\equiv \gamma_{\perp 0}$. For a stationary plasma the only relativistic motion is in the $\perp$ direction. Therefore, for large $a_0$ the growth rate scales $a_0^{-1/3}$ (recall that $M_0=1$ in this case) which is a less rapid fall off compared to the $a_0^{-2/3}$ scaling for the pair plasma. The physics behind the scaling with $a_0$ in the stationary plasma is easy to understand \cite{decker1996evolution}.  There is a $\gamma_{\perp 0}^{-1}$ correction to the rest mass, fluid velocity,  and the ponderomotive force leading to a $\gamma_{\perp 0}^{-3}$ factor in addition to the $a_0^2$ term inside the parenthesis. On the other hand the pair plasma drifts forward as well as oscillates so the $\gamma$ factor is now $(1+a_0^2/2)\equiv \gamma^{(0)}$. Furthermore, the proper density is an invariant, so there is only a single $\gamma^{(0)^{-1}}$ factor. However, there is a Doppler downshift to the backscattered frequency (the $M_0$ factor) which leads to an additional $[\gamma^{(0)}]^{-1}$ factor. Although there are a smaller number of relativstic $\gamma$ factors,  the dependence of $\gamma$ on $a_0$ is stronger for the pair plasma. 



\begin{figure}[H]
     \centering
     \begin{subfigure}[b]{0.49\textwidth}
         \centering
         \includegraphics[width=\textwidth]{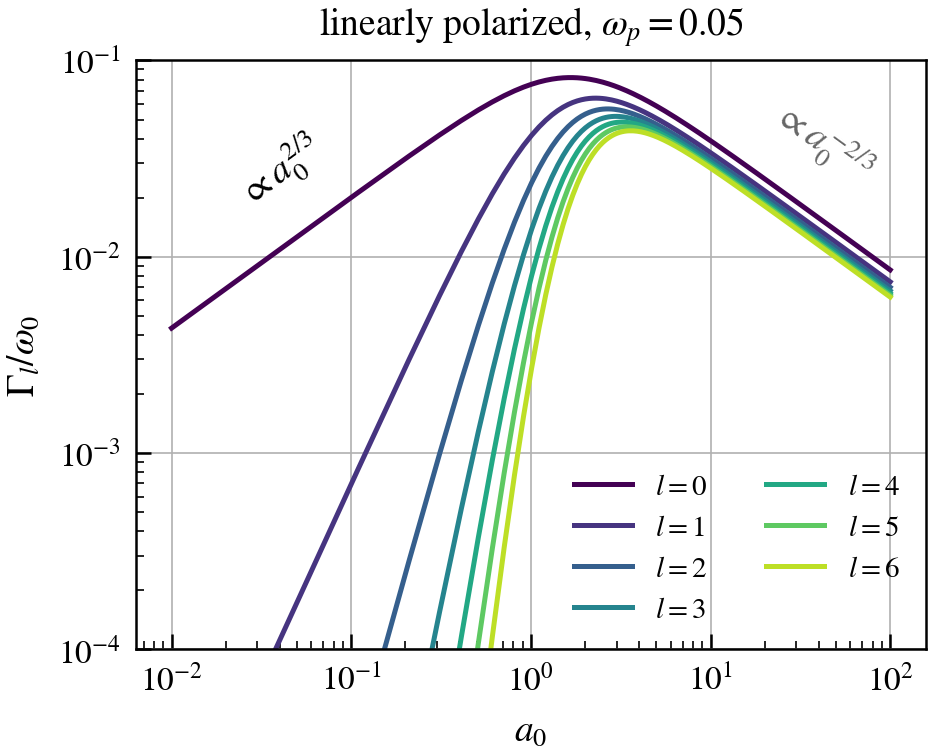}
     \end{subfigure}
     \begin{subfigure}[b]{0.49\textwidth}
         \centering
         \includegraphics[width=\textwidth]{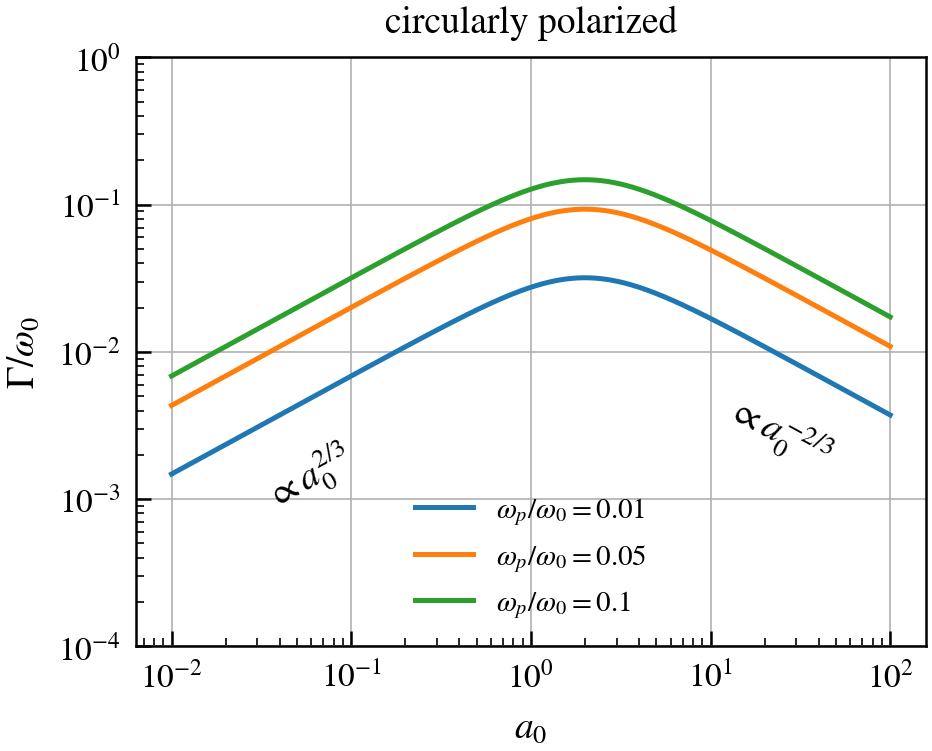}
     \end{subfigure}
        \caption{Plots of growth rate as a function of pump strength $a_0$ from \ref{eq:growth-rate} for linearly polarized (left) and circularly polarized pumps (right). The left plot depicts the growth of each harmonic along with the fundamental mode for $\omega_p/\omega_0 = 0.05$. Note that the harmonics only become significant for $a_0$ near unity.}
        \label{fig:disp-plot}
\end{figure}

To illustrate the dependence on the growth rate on $a_0$ for both LP and CP, in \ref{fig:disp-plot} we plot the solutions to  the above dispersion relation for the 
LP case for $l = 0-6$ 
(left) and the 
CP case (right).

For the $a_0<1$ limit, there is a single parameter that determines $\Gamma/\omega_0$, namely 
$\epsilon_p \equiv a_0\omega_p/\omega_0$. However, as $\frac{\omega_p}{\omega_0}$ increases then the dispersion properties of the pair plasma cannot be neglected and $v_{\phi}$ no longer is exactly c. In this situation the incoming EM wave is no longer  purely a function of $z-ct$ and the above theory no longer holds. It is difficult to undertake a general theory for LP lasers for arbitrary pump phase velocity, however, it can be done for CP and this is in the Appendix. 

Furthermore, as $a_0$ increases beyond unity, then the relevant parameter that determines the normalized growth rate is now $[a_0\omega_0/\omega_p]^{-1}]$ which is always much less than unity. However, the parameter, $\epsilon_p$, is still very important for large $a_0$ for a different reason, local pump depletion or etching \cite{decker1996evolution}. As the EM wave enters the pair plasma it accelerates the pairs forward. Each pair has an energy $\gamma^{(0)}mc^2$. In addition, the proper density is an invariant, so the density is compressed to $n_0 \gamma^{(0)}$ (recall that $n_0$ is the proper density). There is thus a large local kinetic energy density for the pair plasma. This energy must come from the incident EM wave. Each laser cycle thus shoots pairs forward, so when the energy density of the pairs exceeds the energy density of the incoming EM wave then the wave gradually etches backwards 
it cannot propagate into the plasma. Equating the two quantities gives,
 \begin{equation}
    2 n_0 \gamma^{(0)}(\gamma^{(0)}-1)mc^2 = 2 n_0 (1+a_0^2/2)(a_0^2/2) mc^2=\frac{E_0^2}{8\pi}.
    \label{eq:etching_1}
\end{equation}
Noting that $\frac{eE_0}{mc\omega_0}\equiv a_0$ and defining complete etching to occur when the energy density of the pairs exceeds that of the EM wave leads to the condition, $\epsilon_p>1$, for full etching under the assumption that  $a_0>>1$. Under these conditions, there is no reflection of the EM wave. It is fully absorbed by the pair plasma.

Last, we point out a related issue. For low density, where the group velocity of the EM wave is very close to the speed of light, then it moves faster than the pair plasma that is accelerated forward. In this situation the EM pulse will interact with a pair plasma of an ever increasing thickess  given by $\Delta=(1-v_0/c) L_p$ where $L_p$ is the propagation distance of the EM pulse into the pair plasma and $v_0/c\equiv \frac{a_0^2/2}{1+a_0^2/2}$ is the average axial velocity of the pair plasma. It thus follows that $\Delta=L_p/\gamma^{(0)}$ which is the Lorentz contraction of the propagation distance. 

At higher density the EM pulse does not move at c rather at near the group velocity of the EM pulse, $v_g/c\approx 1-\frac{1}{2}\frac{\omega_p^2}{\omega_0^2}$. In this case $\Delta=\left(\frac{1}{1+a_0^2/2}-\frac{1}{2}\frac{\omega_p^2}{\omega_0^2}\right )L_p$ which for large $a_0$ becomes negative (the EM pulse can never catch up with the drifting pair plasma and be reflected) when $\epsilon_p >2$. Thus, for large $\epsilon_p$ the EM pulse cannot undergo stimulated scattering and it is instead, suddenly or gradually absorbed. 

\section{Simulation Results}
We verify aspects of the above theory and arguments using the fully relativistic particle-in-cell code OSIRIS. We perform two types of numerical experiments. The first utilize periodic boundary conditions 
with the pump 
initialized {\it inside} the plasma with the 
a density and fluid velocity 
determined from the equilibrium 
of Eqns. (\ref{eq:equil1})-(\ref{eq:equil3}) for a pair plasma that started at rest in front of the pump.
These simulations permit 
numerical validation of the  temporal growth of the instability predicted by the theory. We call this the temporal problem. The second class of numerical experiments represent a more realistic initial condition, where the pump is launched from the left boundary into a stationary pair plasma. We call this the spatial problem.

\subsection{Temporal Problem}
\begin{figure}[H]
\centering
     \begin{subfigure}[b]{0.49\textwidth}
         \includegraphics[width=\textwidth]{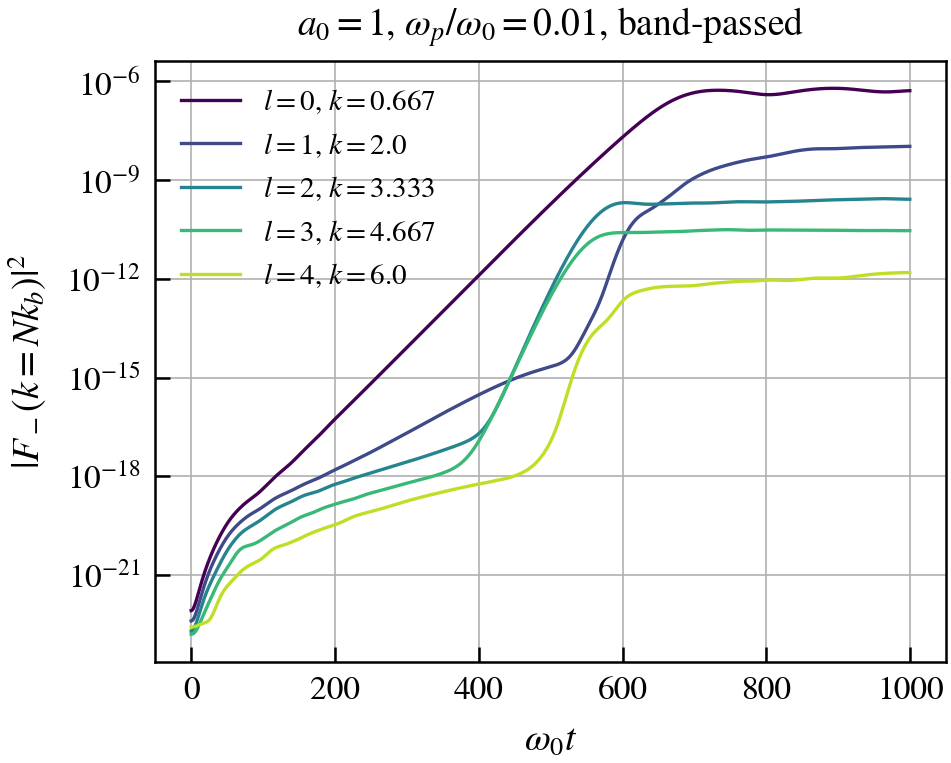}
         \caption{}
         \label{fig:harmonic-growth-a0-1}
     \end{subfigure}
\centering
     \begin{subfigure}[b]{0.49\textwidth}
         \includegraphics[width=\textwidth]{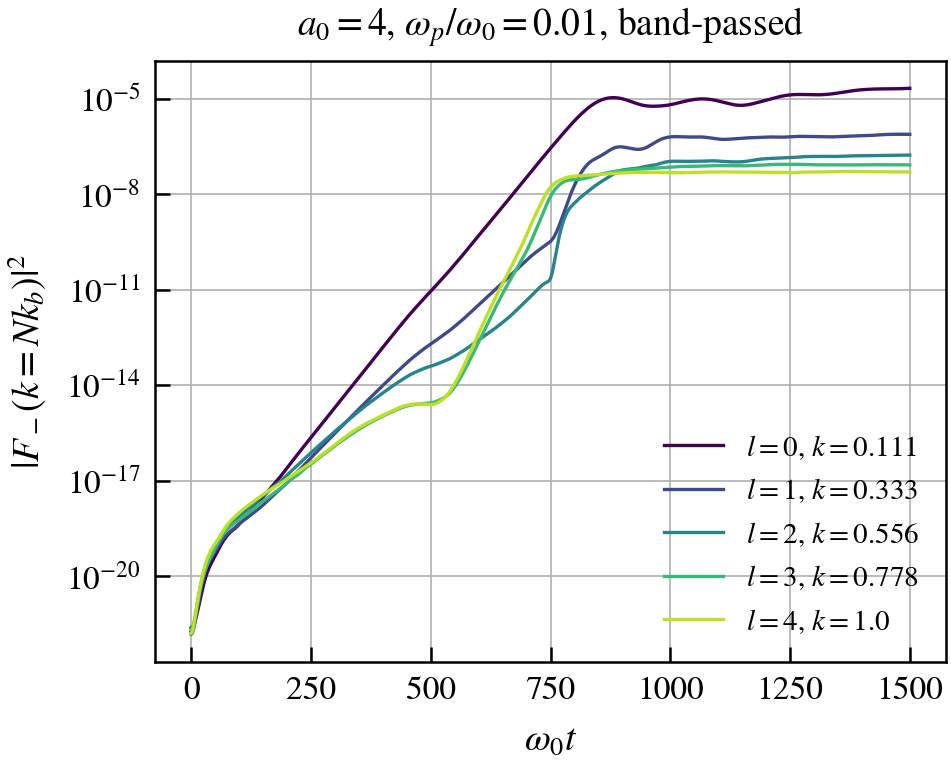}
         \caption{}
         \label{fig:harmonic-growth-a0-4}
     \end{subfigure}
     
\begin{subfigure}[b]{0.49\textwidth}
         \centering
         \includegraphics[width=\textwidth]{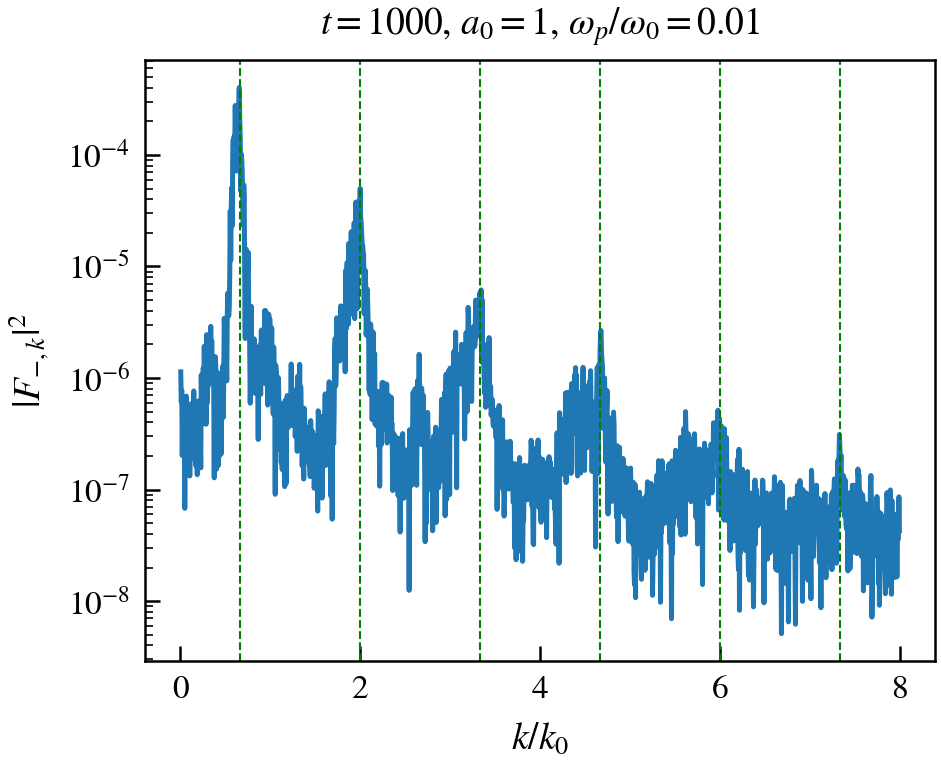}
         \caption{}
\end{subfigure}
\begin{subfigure}[b]{0.49\textwidth}
         \centering
         \includegraphics[width=\textwidth]{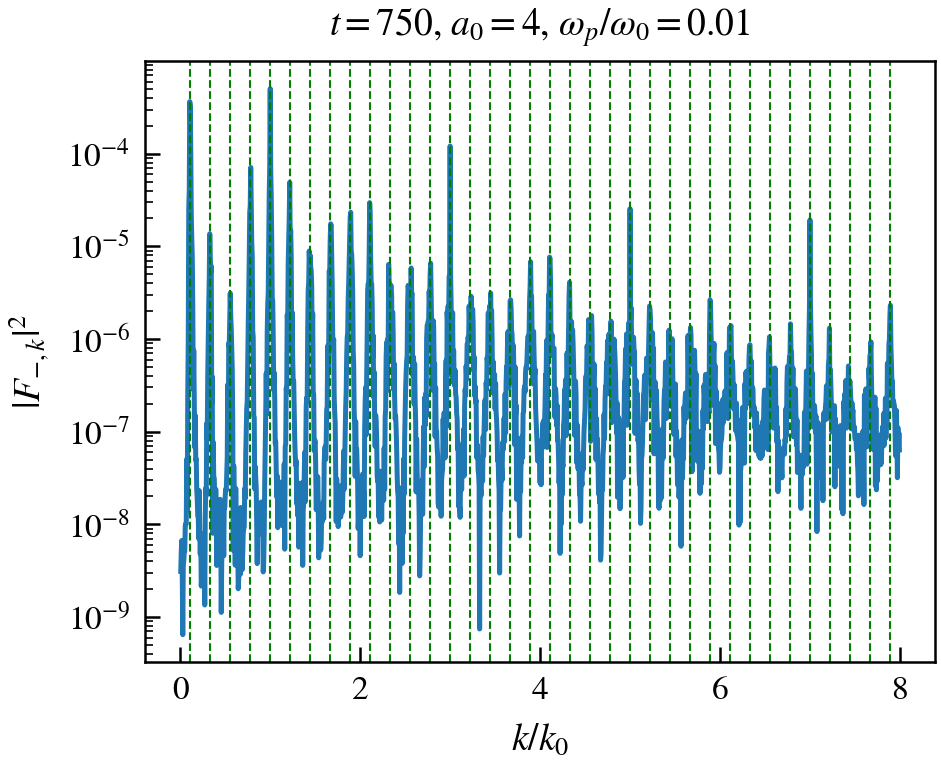}
         \caption{}
\end{subfigure}
\caption{Exponential growth of the Fourier amplitude ($|F_{-}|^2$) of the backscattered harmonics for $l = 0,...,4$ as a function of time for different pump strengths: (a) $a_0 = 1$ and (b) $a_0=4$; $\omega_p/\omega_0 = 0.01$ in both cases. The corresponding k-spectrum of the backscattered energy at late times, after the instability has saturated, is presented in panels (c) and (d), revealing the harmonic structure of the backscattered light. Dashed lines in (c) and (d) are drawn at $k_l = -(2l+1)k_0/(1+a_0^2/2)$}.
\label{fig:kt-lineouts}
\end{figure}

The simulation lengths for the temporal problem are 60-180 laser wavelengths
with
6k-18k cells and
the number of particles per cell ranging from 64-128. The ratio of the plasma density to the frequency of the incident EM wave is
plasma density is 
$\omega_p = 0.01$. Simulations were done for both linearly and circularly polarized laser pumps for $a_0$ ranging from $2^{-4}-2^3$.  As seen in Fig. [\ref{fig:kt-lineouts}], which plots the growth of $F_- \equiv (E_2 - B_3)$ in $k$-space, each mode grows for a duration of $\omega_0 t\sim 10^3$ before it saturates. Note that at early times, the growth rate is noticeably 
lower for 
the 
higher harmonic numbers, $l$. The amplitudes saturate near 
$\omega_0 t\sim 600$. 

In the bottom row of Fig. [\ref{fig:kt-lineouts}], we plot the wave number spectrum of the backward-propagating component $F_-$ of the EM field at a time after saturation, $\omega_0 t\sim 750$ for the corresponding EM wave amplitudes.

For each wave amplitude, the most prominent mode is the fundamental harmonic that is downshifted in frequency by a factor of $M_0=1 / (1 + a_0^2/2)$ due to the nonlinear relativistic drift of the plasma particles. 

\begin{figure}[H]
     \centering
     \begin{subfigure}[b]{0.32\textwidth}
         \includegraphics[width=\textwidth]{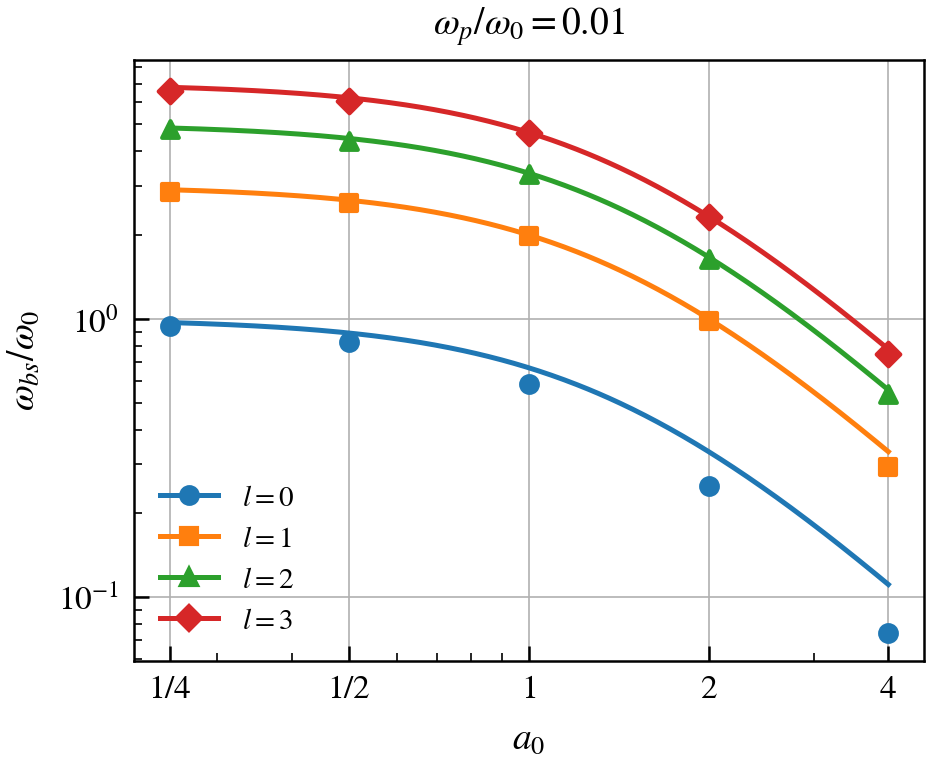}
         \caption{}
         \label{fig:doppler-plot}
     \end{subfigure}
     \begin{subfigure}[b]{0.32\textwidth}
         \centering
         \includegraphics[width=\textwidth]{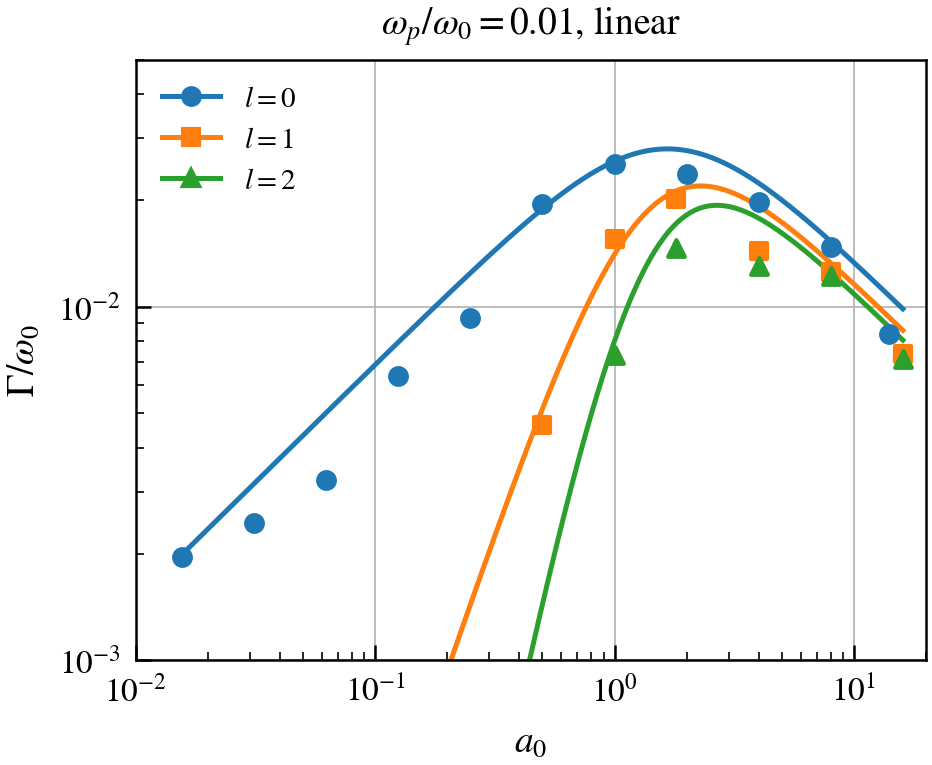}
         \caption{}
         \label{fig:growth-rate-lin}
     \end{subfigure}
    \begin{subfigure}[b]{0.32\textwidth}
         \centering
         \includegraphics[width=\textwidth]{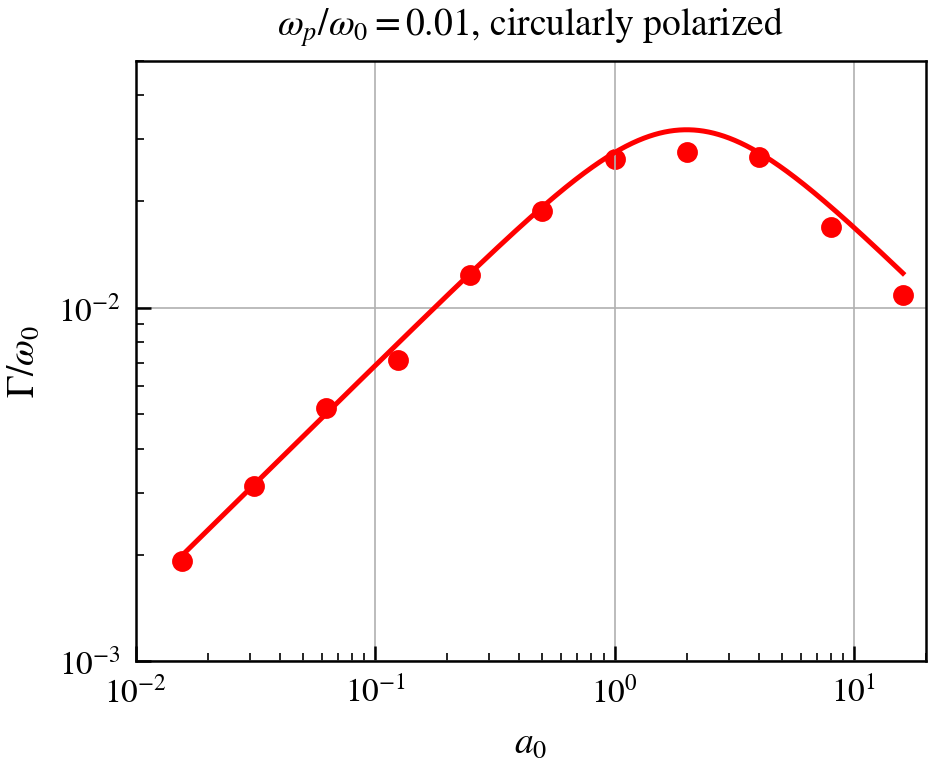}
         \caption{}
         \label{fig:growth-rate-circ}
     \end{subfigure}
     
        \caption{a) Plot of the $l = 0,..,3$ frequency mode vs pump strength $a_0$. Frequencies were measured by extracting the maximum spectral signal of the backscattered light wave $F_- = E_2 - B_3$. Plots of the growth rate as a function of $a_0$ for b) linearly and c) cirularly polarized pumps. The blue and orange solid curves denote the predicted growth rates as predicted by the $l = 0$ FEL dispersion curve and the linear theory. }
        \label{fig:doppler-harmonic-growth}
\end{figure}

We can extract this frequency from the PIC data as well as the frequency at all the harmonics by measuring the corresponding spectral peaks near each shifted frequency. We plot the PIC backscattered frequencies vs theory for the $l = 0$ and $l = 1$ modes in the linearly polarized case in \ref{fig:doppler-plot}.
 The leftmost plot shows the measured location of each harmonic in $\omega$-space vs $a_0$. We measure the location of each harmonic by fitting a gaussian to each peak in the power spectrum obtained from a lineout at the $\omega = k$ line in the dispersion. There is good agreement at small $a_0$, for each harmonic, however there is noticeable deviation of the fundamental at large $a_0$.
 
We can extract the numerical growth rate at each harmonic  from the temporal evolution of the electron density at $k = (2l + 1)M_0 k_0$. At various $a_0$ we observe distinct regions of exponential growth in the $S_{1,k},|F_{-,k}|^2,|n_{e,k}|^2$ plots and apply a linear regression to the semilog of the data, with the growth rate being half of the slope of the best fit line. The exponential growth of each harmonic is shown in Figs. (\ref{fig:harmonic-growth-a0-1}) and (\ref{fig:harmonic-growth-a0-4}).
We have carried out this procedure to extract the growth rates for a range of wave amplitudes,  $a_0 = 2^{-4}-2^3$, and plotted the growth rate for linearly and circularly polarized cases in Figs. (\ref{fig:growth-rate-lin}) and (\ref{fig:growth-rate-circ}). The fully nonlinear growth rate as predicted by Eqn. (\ref{eq:growth-rate}) for linearly and circularly polarized modes are shown as the blue and red curves. Note the characteristic bending of the growth rate at large $a_0$ observed in the data that is not accounted for by the linear theory. For further numerical validation, we also plot the growth rates of the harmonics along with the fundamental for the LP case. 

\subsubsection{Nonlinear effects}
\begin{figure}[H]
     \centering
     \begin{subfigure}[b]{\textwidth}
         \centering
         \includegraphics[width=\textwidth]{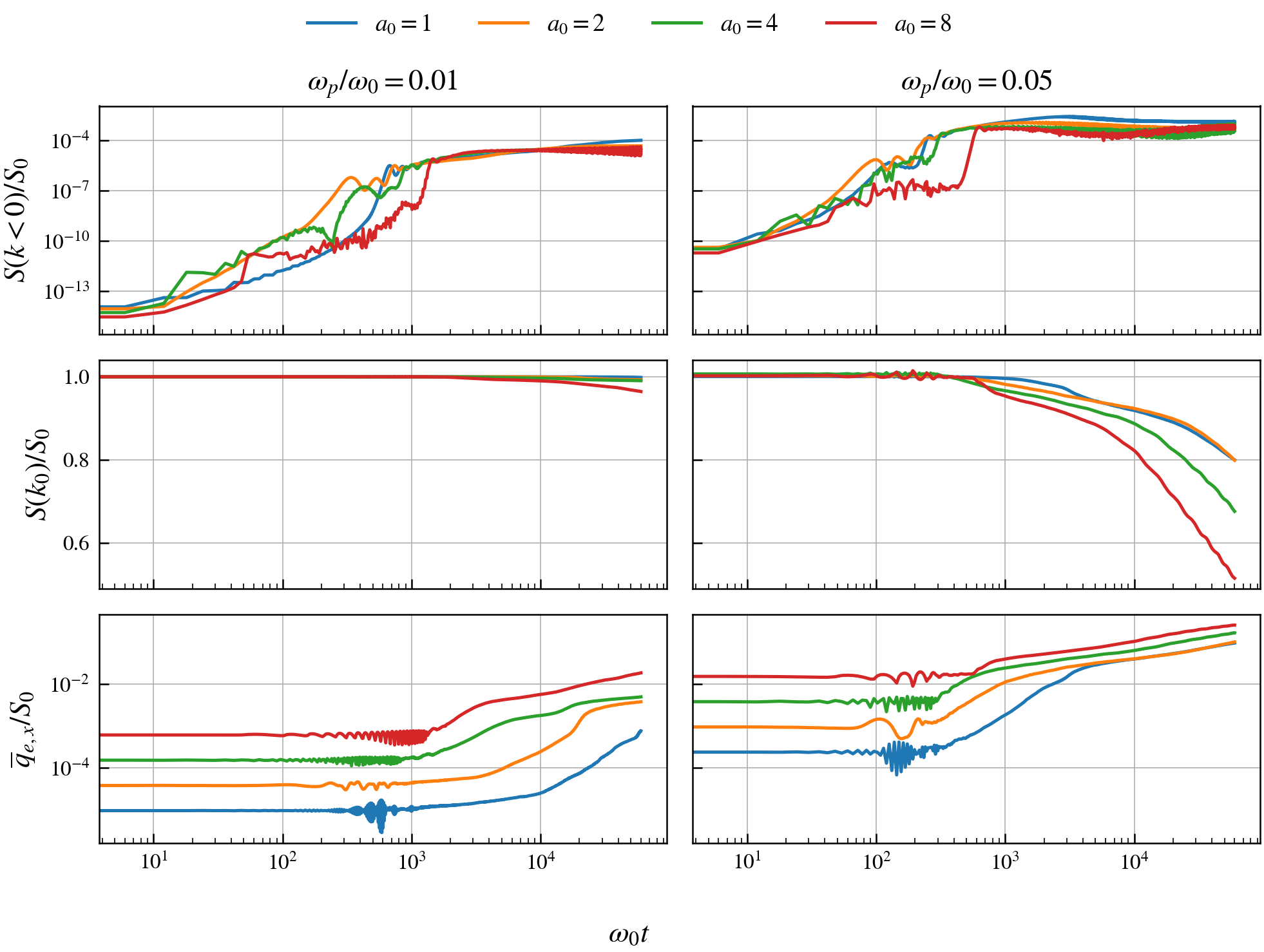}
     \end{subfigure}
        \caption{Plots of the Poynting fluxes of the backscattered harmonics (top row), the incident EM wave (middle row), and of the electron energy flux (bottom row). The left (right) column corresponds to $\omega_p/\omega_0 = 0.01~(0.05)$.
        The colors correspond to different values of $a_0$ which vary from 1 to 8 in powers of 2. }
        \label{fig:energy-flux}
\end{figure}

For these periodic simulations it is illustrative to examine the nonlinear behavior of the instability. 
This process is illustrated in Fig. \ref{fig:energy-flux}
The top plot shows the total backscattered Poynting flux summed over all $k <0$ modes, the middle shows the Poynting flux filtered near the pump at $k_0$ and the bottom plot shows the spatial averaged electron energy flux vs time. The complex Poynting fluxes were extracted using the method described in \cite{Ghosh2022}. The linear regime of the instability is readily seen to occur in the time range $t   = 10^2-10^3$, which is associated with a rapid increase in the backscatter energy flux. From $t =10^3$ onwards, we have entered the nonlinear regime. In this region, there is pump depletion of a few percent for $\omega_p = 0.01$ and a few tens of percent for $\omega_p = 0.05$ for the durations considered. 
This energy does not go into the reflected modes, but instead goes into heating the particles, which is readily scene by the steady rise in the bottom plot which occurs at the same time as the pump depletion and the saturation of the reflection. 
One possibility for the mechanism for the onset of pair plasma heating 
is a form of stochastic acceleration in which the particles orbit chaotically due to the counterpropagating light waves\cite{mendonca}. This process has been observed in simulations of SRS backscatter of electron-ion plasmas \cite{ForslundHeating}.

\subsection{Spatial Problem}
\begin{figure}[H]
     \centering
     \begin{subfigure}[b]{0.49\textwidth}
         \centering
         \includegraphics[width=\textwidth]{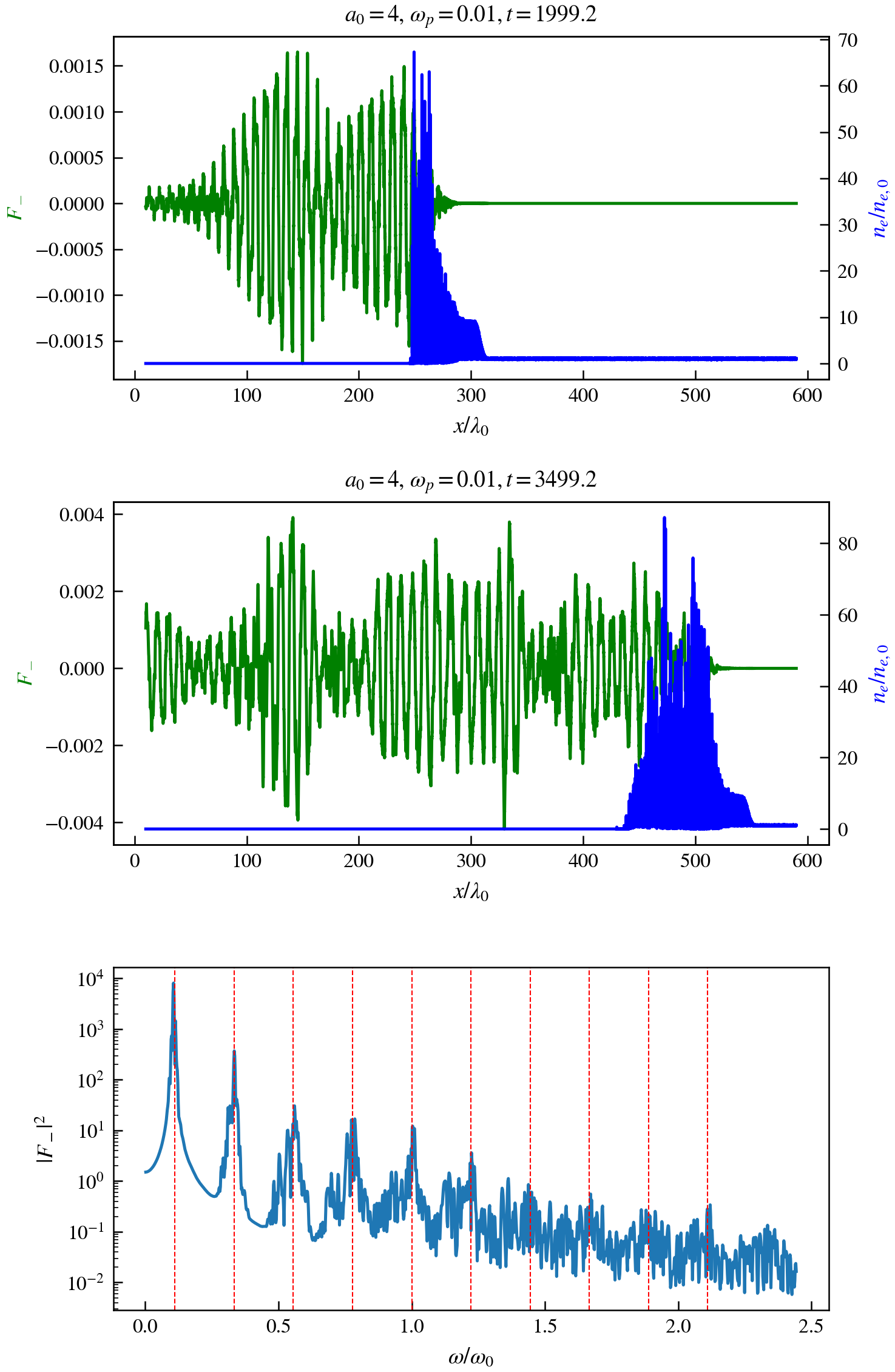}
         \caption{}
     \end{subfigure}
    \begin{subfigure}[b]{0.49\textwidth}
         \centering
         \includegraphics[width=\textwidth]{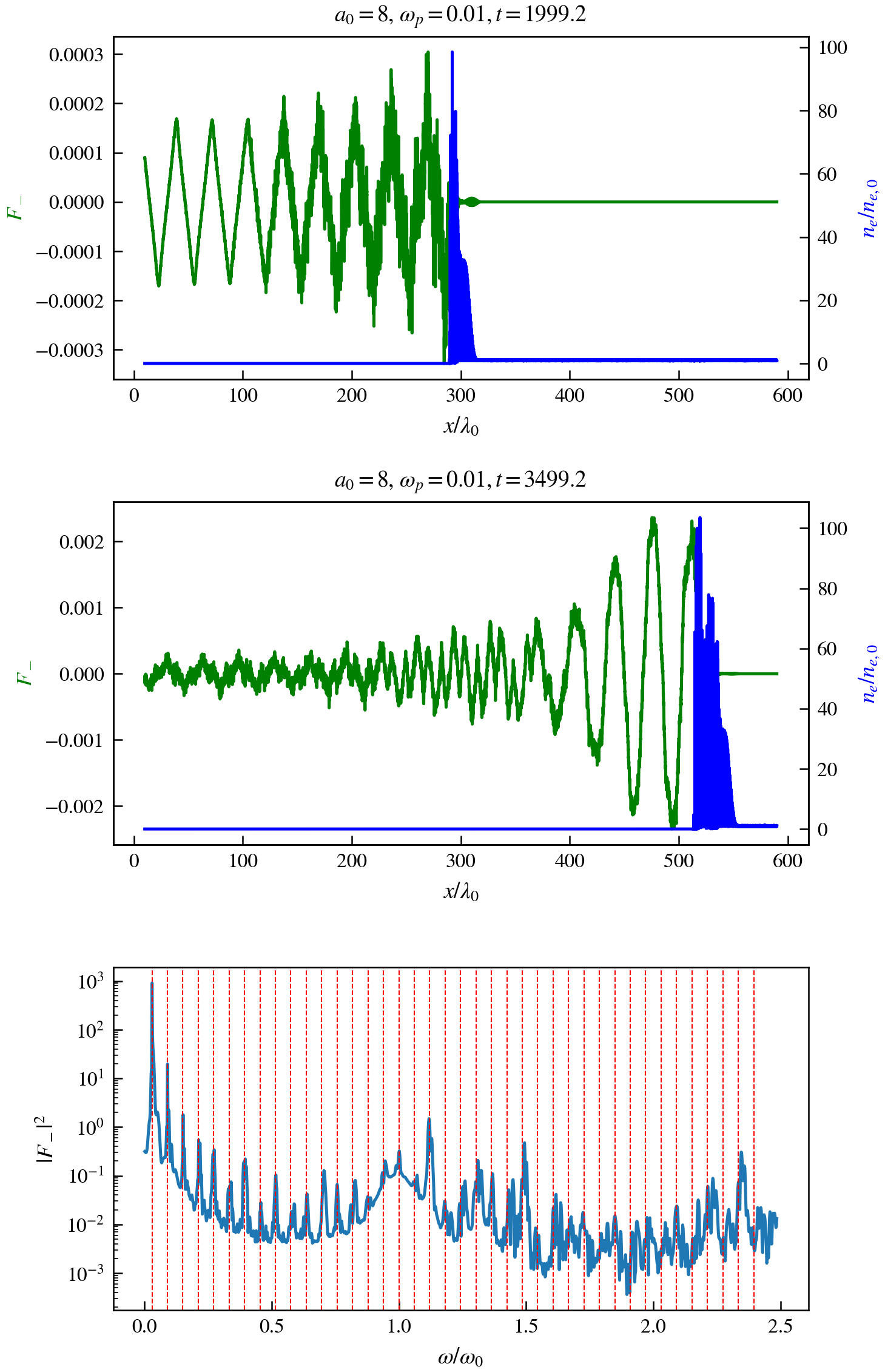}
         \caption{}
     \end{subfigure}
        \caption{Spatial plots of the electron density $n_e/n_{e,0}$ and leftward going field $F_-$ at $t = 2000c/\omega_0$ for $a_0 = 4$ and $a_0 = 8$ for linearly polarized pump on top. The flat part of the density at the front of the laser is precisely the equilibrium profile $\rho^{(0)} = 1 + {a_{(0)}}^2/2$ which then goes unstable further into the pulse. Spectral analysis of reflected light on bottom. The signal between $10\lambda_0$ and $50\lambda_0$ was fourier transformed in time from $t = 1000c/\omega_0$ to $3000c/\omega_0$ to avoid interference from the primary pulse.}
        \label{fig:spatial}
\end{figure}

While our temporal simulations provide numerical consistency with our theory, a simulation investigating the behavior of the pump when launched from a boundary is more physical in the context of the propagation of fast radio bursts. For these simulations, we use a box size of 600 laser wavelengths, with 60k-180k cells and 64-128 particles per cell. Maintaining the same resolution as the temporal simulations
is important here, as the odd-harmonics require at least 20 cells per wavelength in order to be fully resolved. The antenna module of Osiris is used to launch an electromagnetic wave of $a_0 = 1,2,4$ from the left boundary. The rise time was set to $100c/\omega_0$ in order for the density to achieve the equilibrium form. Evidence of the backscatter can be readily seen in the leftward going EM field component $F_-$. Plots of $F_-$ along with the electron density at various $a_0$ are shown\ref{fig:spatial}.
In each case, the density at the front of the EM wave follows the equilibrium form at the head of the laser pulse before becoming unstable. There is strong evidence of backscatter past this point. A spectral analysis of the reflected signal is presented in Fig. (\ref{fig:spatial}) in the bottom plot which reveals the locations of the fundamental and odd-harmonic generation in close agreement with theory. It is evident that the bacskcattered fundamental mode has increasing wavelength with increasing $a_0$, which is a feature of the relativistic Doppler shift. For the $a_0 = 4$ case, visual inspection reveals that there around 10 backscattered wavelengths for every 100 laser wavelengths, so that the backscattered frequency is $\sim 0.1\omega_0$. The predicted frequency of the fundamental is $\omega_0/9$. Similarly there are 3 backscattered wavelengths for every 100 laser wavelengths for $a_0 = 8$, yielding a backscattered frequency of $\sim 0.03\omega_0$ and for comparison the predicted frequency is at $\omega_0 /33$. Another important feature is that as the pulse propagates, the instability region gets progressively longer as plasma piles up at the compression front. This results in the region of spatial gain scaling as $L_p/\gamma^{(0)}$ such the amount of gain increases as the propagation length increases. 
This is illustrated in the middle row of Fig. (\ref{fig:spatial}) which shows the amplification of the backscatter for $a_0 =4,8$ when the pump has reached $300\lambda_0$ and $550\lambda_0$ (near the 
end of the box). For $a_0=4$ it appears that backscatter has almost saturated at $300\lambda_0$ while for $a_0=8$ the backscatted signal has continued to grow and it has increased by an factor of 10 by end of the simulation. Although the amplitude of the reflected energy is low, it is worth noting that the reflected normalized vector potential is actually large. The backscattered electric field is $\approx .05$ that of the pump. However, the signal frequency is .03 of the pump so the normalized vector potential of the reflected wave is slightly larger than that of the pump.

\begin{figure}[H]
     \centering
     \begin{subfigure}[b]{0.8\textwidth}
         \centering
         \includegraphics[width=\textwidth]{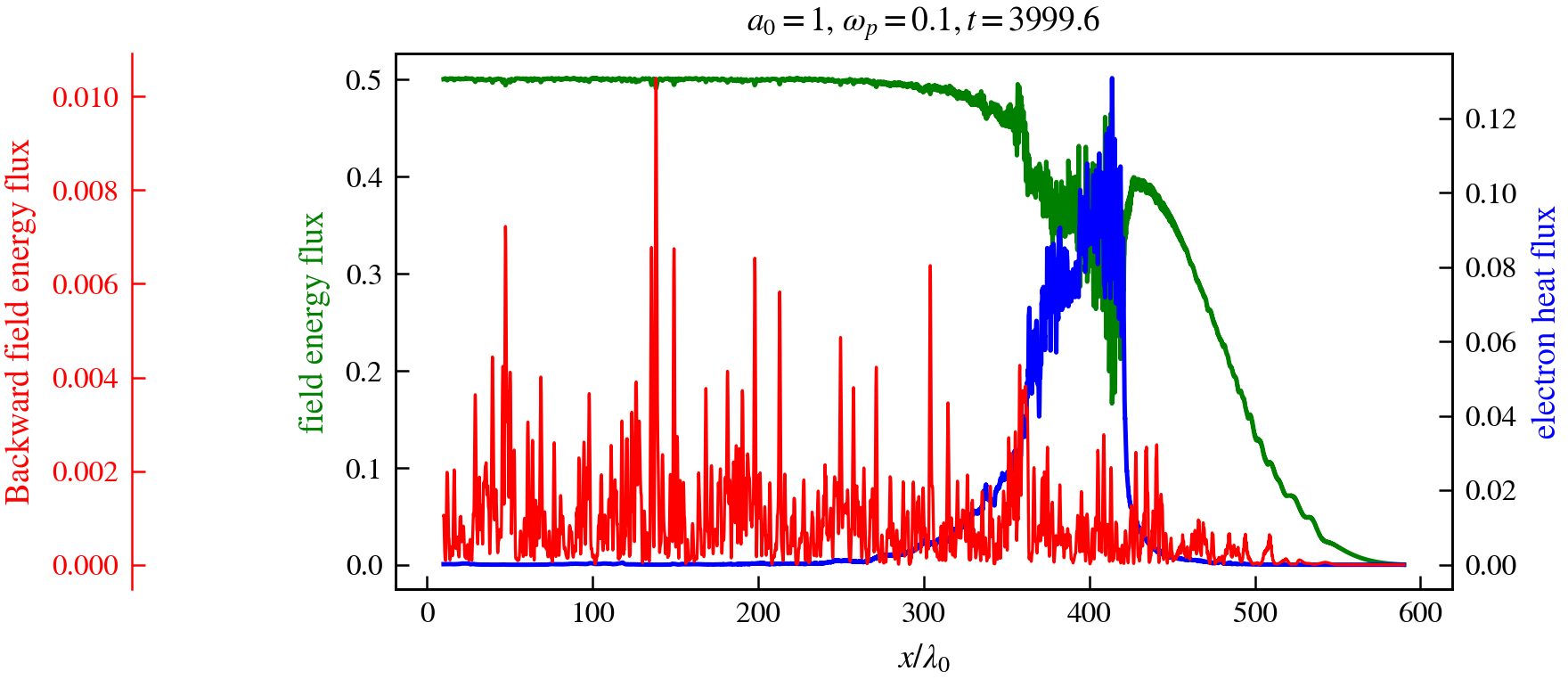}
         \caption{}
     \end{subfigure}
     
    \begin{subfigure}[b]{0.8\textwidth}
         \centering
         \includegraphics[width=\textwidth]{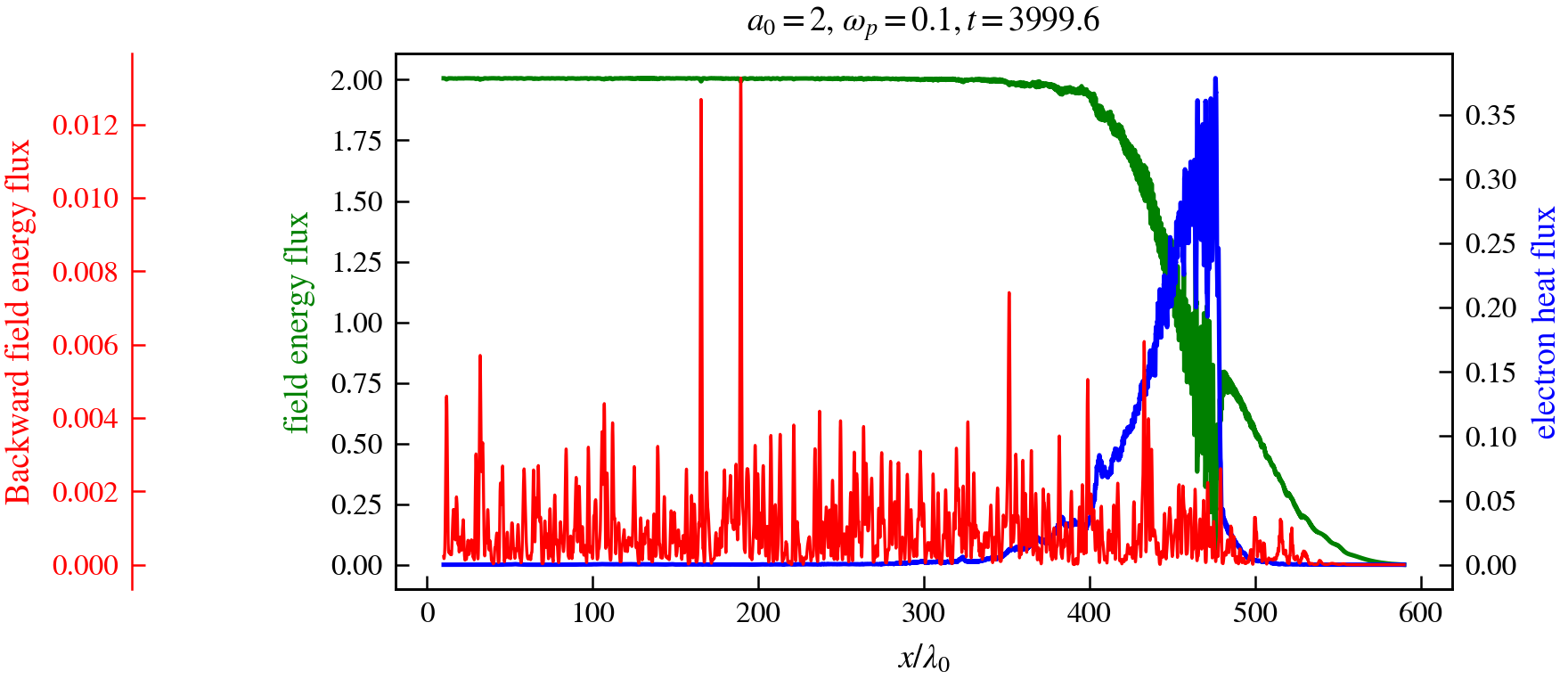}
         \caption{}
     \end{subfigure}
     \begin{subfigure}[b]{0.8\textwidth}
         \centering
         \includegraphics[width=\textwidth]{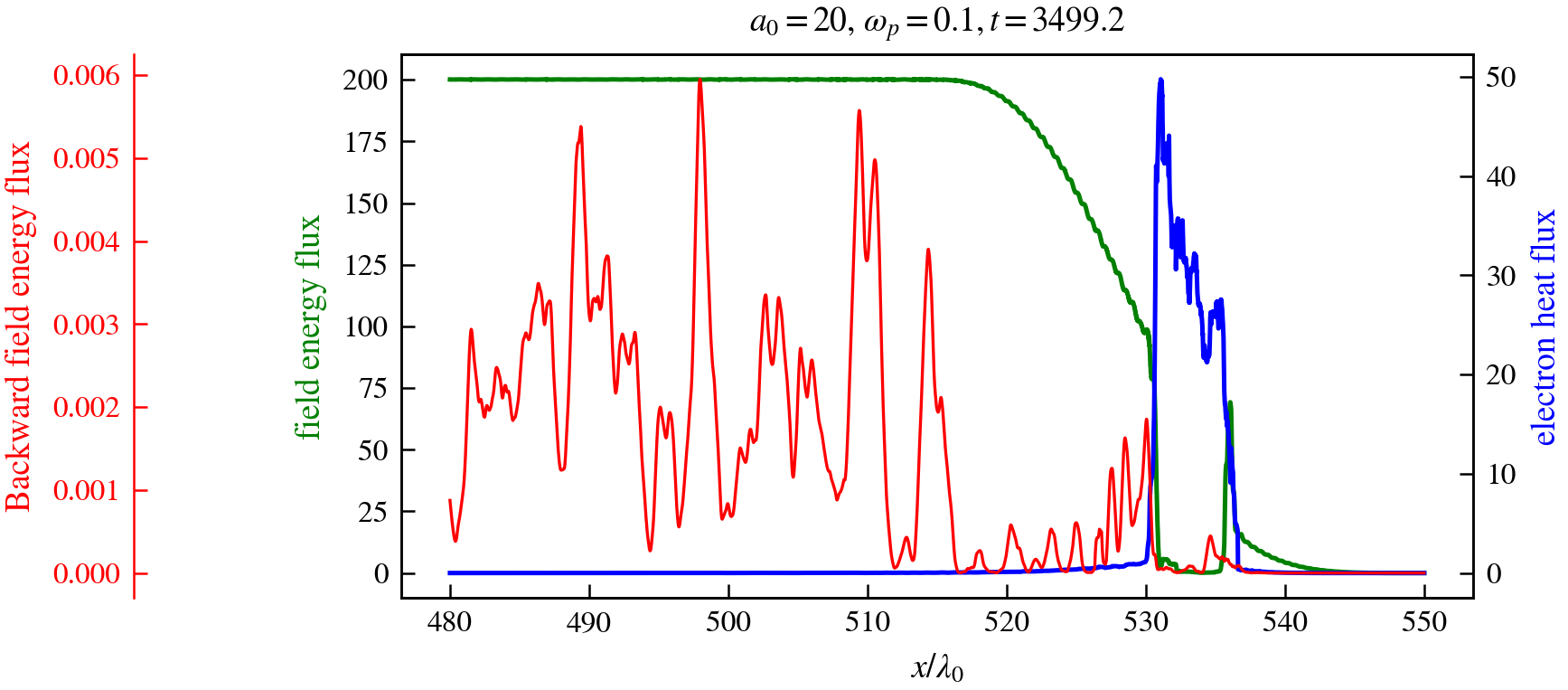}
         \caption{}
     \end{subfigure}
        \caption{Smoothed energy fluxes for particles and fields at $a_0 = 1,2$ and $\omega_p = 0.1$.}
        \label{fig:absorption}
\end{figure}

{\indent} One might wonder if a greater reflected signal is observed at higher densities where the plasma response becomes important. It turns out that for higher $a_0 \omega_p / \omega_0$, much of the incident laser energy is {\it absorbed} by the plasma rather than reflected, with the reflected signal accounting for very little of the incident energy flux. This is shown in Fig. (\ref{fig:absorption}) which depicts the spatial problem for $\omega_p = 0.1$ and $a_0 = 1,2,20$. In these plots the electron heat flux in blue is shown alongside the total Poynting flux $|F_+|^2-|F_-|^2$ in green and the backward Poynting flux $|F_-|^2$ in red. For illustrative purposed the rise time was increased to $t_{rise}=250\lambda_0/c$ for the top two plots, and decreased back to $t_{rise}=40\lambda_0/c$ for the bottom plot.  Note that the cause of the dip at the head of the pulse is due to a spike in the heat flux, rather than to a spike in the reflected signal, which is at most a few percent of the pump energy. Furthermore, for $a_0\omega_p/\omega_0$ near unity as in the bottom plot, the dip in the laser envelope extends all the way to 0, in which case the laser is completely absorbed by the plasma and no more energy can penetrate it. The absorption does not happen when the laser attains its peak energy, but rather when the light Poynting flux equals the energy flux of the leptons when they are accelerated to the drift velocity, which occurs near $\epsilon_p = 1$ for $a_0>>1$. 

\subsection{Finite Temperature effects}
For the temporal and spatial results presented above the plasma was initialized under the conditions that the pair plasma was cold in front of the EM wave. This was done chiefly to permit comparison to theory, but it also simplifies the interpretation of the results.  However, pair plasmas near a magnetar are expected to have non zero temperatures. A natural question would be to determine how this effects the growth rate and the harmonic content. In this section, we provide simulation results for both the temporal and spatial problems for non-zero but equal temperatures for the pair plasma.

\subsubsection{Initialization}
 For the temporal problem, we must determine how to map the distribution function for the pair plasma in equilibrium with a finite initial temperature and no net drifts to the distribution function after the EM wave has propagated through it.
 To determine the appropriate distribution function we rely on Jean's theorem \cite{Jeans1915} which states that any distribution function, F,  constructed from constants of the motion satisfies the Vlasov equation. When the EM wave has a phase velocity equal to c then there are three constants of motion, $\gamma - u_z$, and  two more from the components of $\mathbf{u}_\perp - \mathbf{a}_\perp$. 
 To determine the correct functional dependence for F in terms of  these constants we map an initial $F\propto e^{\beta[\gamma-1]}$ where $\beta\equiv m_ec^2/k_BT$ to the values at some later time after the laser has passed through the plasma. 
 For linear polarization it can be shown that the form of the distribution function is:

 \begin{equation}
     F_{\pm}(\mathbf{u},z) \propto e^{-\beta\big[[(\gamma -u_z - 1)^2+(u_x\pm a_0\cos(z))^2 +u_y^2]/[2(\gamma-u_z)]\big]} 
 \end{equation}
 where $+$ and $-$ refer to positrons and electrons. 
To determine the proportionality constant, we integrate $F$ over the proper velocities, 
\begin{equation}
      n_{\pm}(z) = A\int d^3\mathbf{u} F_{\pm}(\mathbf{u},z)
 \end{equation}
 and the constant $A$ is chosen such that the density in front of the laser is $ n_0$. 
 
 It can be shown that integrating over the distribution function within the EM field gives  
 \begin{equation}
     \frac{\hat n_{\pm}(z)}{\hat n_0} = (1 + \frac{a_0(z)^2}
     {2}
     \frac{K_0(\beta)}{K_2(\beta)})\label{eq:density-profile}
 \end{equation}
 where $a_0(z)=\hat a_0(z)cosk_0z$, $\hat a_0$ is an envelope that also moves at c, and  $K_0$ and $K_2$ are modified Bessel functions of the second kind.  Note that the ratio of the Bessel functions is unity for large $\beta$, i.e., small temperatures, and  asymptotes to zero for small $\beta$. Thus, the compression provided by the EM wave decreases as the temperature increases. 
 
\subsubsection{Simulation}
It is straight forward to include kinetic effects for weak (linear) pump strengths and this has appeared in the literature \cite{Ghosh2022}. In this limit, the EM wave's modification to the density is a lower order effect. On the other hand the above fully theory for induced scattering is not easily extended to include kinetic effects.  Instead,  we relay in PIC simulation experiments to understand the effect of temperature on scattering and harmonic generation for LP EM waves. We begin with results for the temporal problems where the observed growth rates can be compared to the cold plasma results and then move on to the spatial problem. 

\begin{figure}[H]
     \centering
     \begin{subfigure}[b]{1.0\textwidth}
         \centering
         \includegraphics[width=\textwidth]{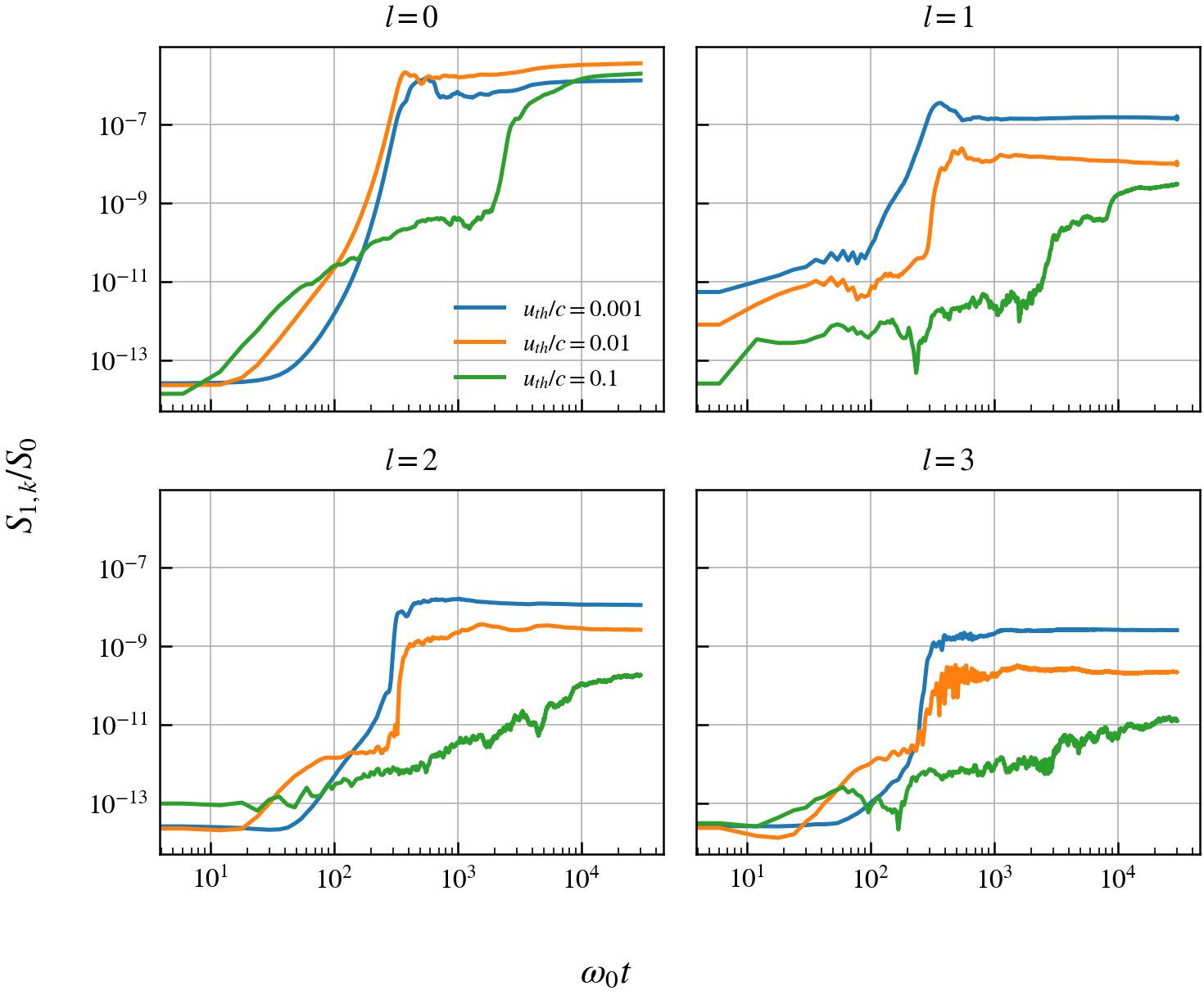}
     \end{subfigure}
    \caption{Poynting flux for finite temperature temporal problem for $a_0 = 2$ and $u_{th}/c = 0.001,...,0.1$.}
     \label{fig:vt-harmonics}
\end{figure}

\begin{figure}[H]
     \centering
     \begin{subfigure}[b]{0.5\textwidth}
         \centering
         \includegraphics[width=\textwidth]{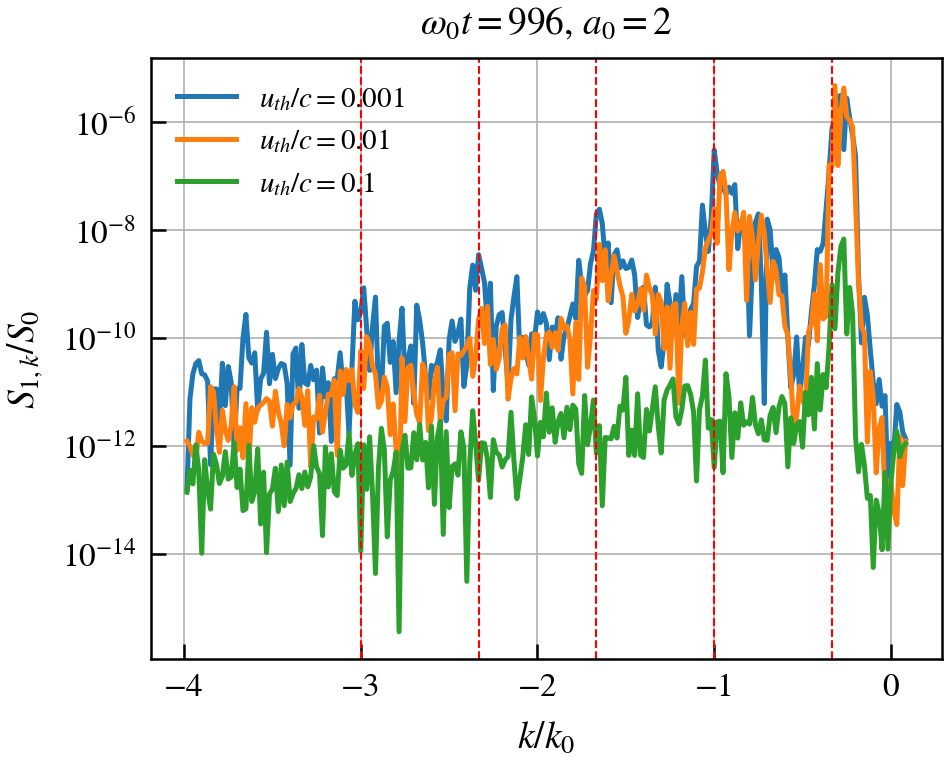}
     \end{subfigure}
    \caption{Lineout for poynting flux for $u_{th}/c = 10^{-3}-10^{-1}$ for $a_0 = 2$}
     \label{fig:vt-harmonics-lineout}
\end{figure}

\paragraph{Temporal Problem} In this section, we perform simulations for plasmas with different temperatures using a periodic box with the same size as for the cold case. The plasma was initialized with a distribution of the form give by Eqn (\ref{eq:density-profile}). We used three temperatures which correspond to $5\cdot 10^{-4}, 5\cdot 10^{-2}$ and $5$ keV.  
The evolution of the harmonics for $l = 0,1,2,3$ for the each temperature is shown in fig. \ref{fig:vt-harmonics}. Each plot depicts the growth of the Poynting flux component at a specific harmonic for temperatures ranging from $u_{th}/c =10^{-3}$ to $10^{-1}$. 
In can be seen that the fundamental appears  for each of the  temperatures simulated. Interestingly, although the initial growth rate of the fundamental decreases with increasing temperature at early times, the growth rate actually appears to increase later in time and the fundamental mode saturates near the same value. 
On the other hand, the saturated levels of the  $l > 0$ modes are decrease as the temperature increases. 
still become less prominent at higher temperatures, with the higher harmonics saturating at two orders of magnitude less than the cold case for $T = 10^{-3}$.  Figure \ref{fig:vt-harmonics-lineout} depicts a line out of the Poynting flux after saturation at three different temperatures. 
In this figure, there are noticeable spikes for the colder case at $T=10^{-6}$ while the hot case $T=10^{-2}$ has only the fundamental harmonic as a noticeable feature.

\begin{figure}[H]
     \centering
     \begin{subfigure}[b]{0.49\textwidth}
         \centering
         \includegraphics[width=\textwidth]{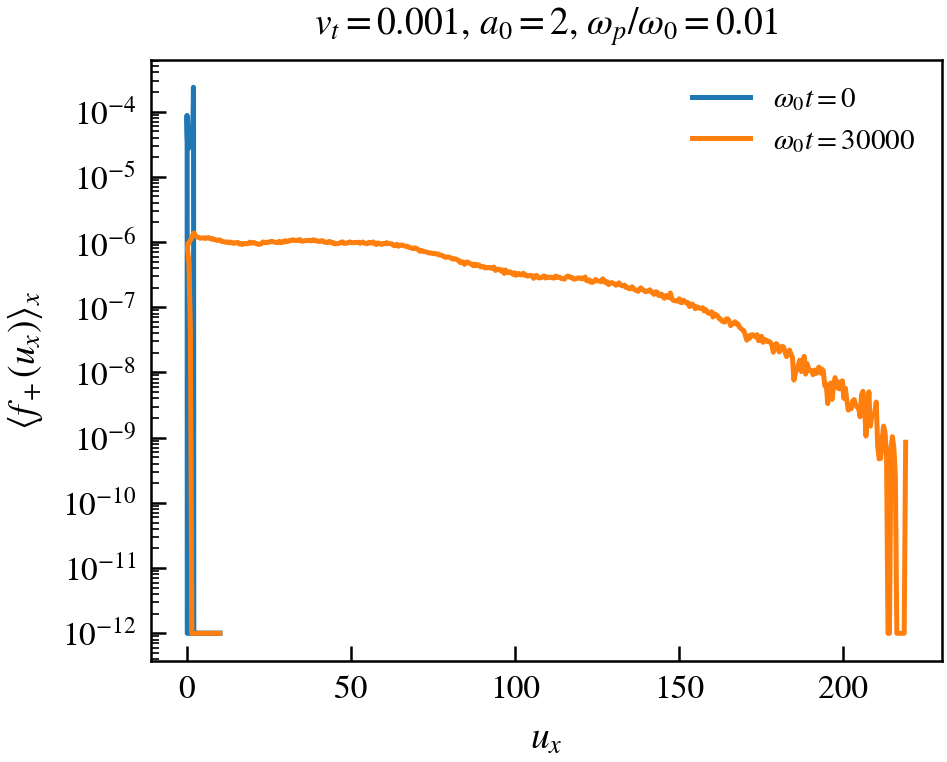}
     \end{subfigure}
    \begin{subfigure}[b]{0.49\textwidth}
         \centering
         \includegraphics[width=\textwidth]{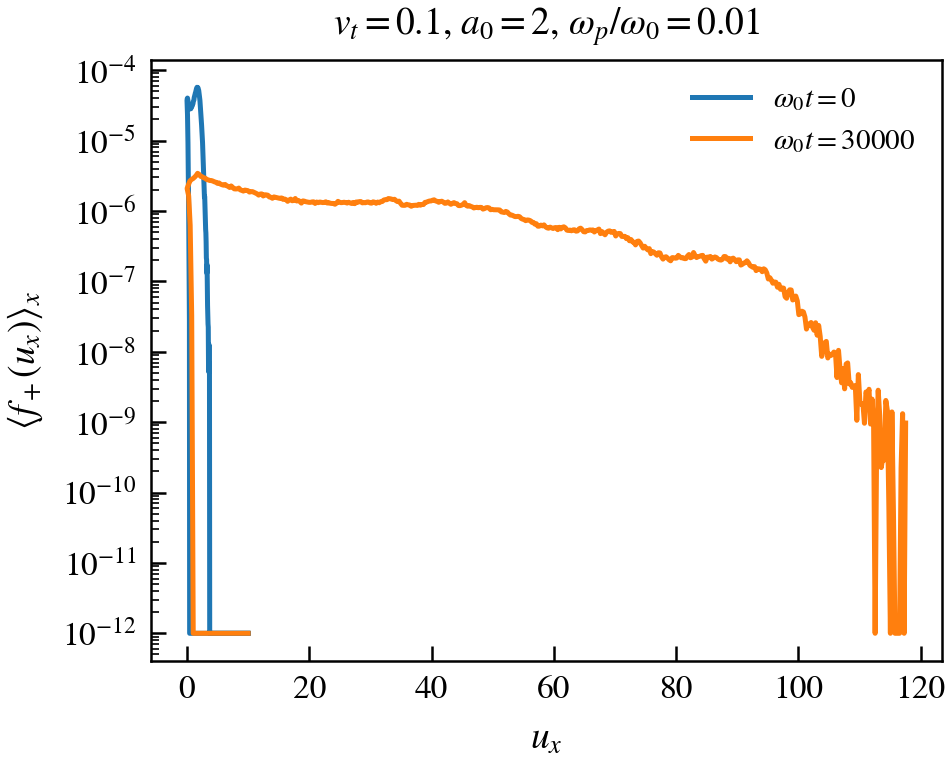}
     \end{subfigure}
        \caption{Initial and final Positron distribution function for temperatures $u_{th}/c = 0.001, 0.1$. Significant heating is observed in both cases with a much greater fraction of particles being resonantly coupled to the pump in the cold case.}
        \label{fig:distribution-plot}
\end{figure}

\begin{figure}[H]
     \centering
     \begin{subfigure}[b]{0.5\textwidth}
         \centering
         \includegraphics[width=\textwidth]{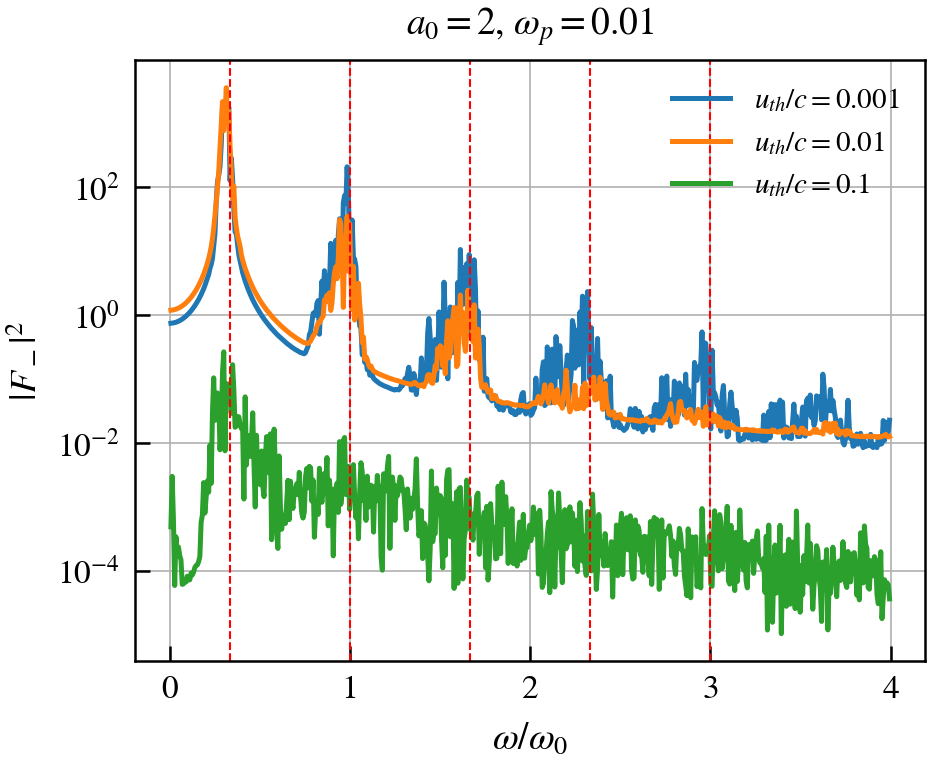}
     \end{subfigure}
    \caption{Backscattered spectra for $u_{th}/c = 0.001,0.01,0.1$. }
     \label{fig:finite-temp-spatial}
\end{figure}

{\indent} We observe signficant heating even in the cold case as the incident EM energy at these amplitudes is much greater than the initial plasma energy. This is illustrated in Fig. (\ref{fig:distribution-plot}) which shows the particle distribution both initially and at a later time for $T = 10^{-6} ,10^{-2}$ note that the distribution function at higher initial temperatures has a 
lower maximum energy 
than the colder case. This is consistent with the ponderomotive force being smaller.
\paragraph{Spatial Problem} We also carried out simulations for finite pair plasma temperatures
for the spatial problem. Results are shown in Fig. (\ref{fig:finite-temp-spatial}). 
We observe a similar phenomenon as for the temporal problem in which the higher harmonics are 
less visible while the fundamental is still present. In the spatial problem there is significantly less energy in both the fundamental and corresponding higher harmonics for increasing temperature. This is potentially due to the slower there being less gain due to the Lorentz contraction to the width of the compressed plasma region.

\section{Application to FRBs}
The above analysis and simulations describe the nonlinear interaction of EM waves with large normalized amplitudes in unmagnetized pair plasmas with small normalized densities. Here, we discuss how these results might be of relevance to  Fast radio bursts (FRBs). FRBs detected from extragalactic sources are thought to originate in or near magnetar magnetospheres \cite{Lyutikov1998,Lyubarsky2008,Sobacchi2022, Sobacchi2024AA, Sobacchi2024PRR}, where violent cataclysmic events launch coherent gigahertz-frequency electromagnetic pulses through a surrounding pair-dominated wind. Deep in the closed-field-line region the strong cyclotron frequency $\omega_B = eB/(m_ec) \gg \omega_0$ makes the unmagnetized model inapplicable, but beyond the light cylinder, the field weakens and the wind transitions toward an approximately unmagnetized pair plasma \cite{Sobacchi2022, Sironi}. In the striped wind \cite{Beloborodov2026} the alternating-polarity magnetic field has a time-averaged value near zero on scales larger than the stripe width, further reducing the effective magnetization. It is in this region that the intense FRB pulse interacts with the pair plasma on the timescale of interest, motivating the present analysis and simulations.

The key dimensionless parameter governing this interaction is the normalized pump amplitude $a_0$. For an FRB of isotropic luminosity $L$ at distance $r$ from the source, the local intensity is $I = L/(4\pi r^2)$, so
\begin{equation}
    a_0 = \frac{e}{m_e \omega_0 c}\sqrt{\frac{2L}{r^2 c}} \approx 2\times 10^4 \left(\frac{L}{10^{42}\,\text{erg s}^{-1}}\right)^{1/2} \!\!\left(\frac{10^{9}\,\text{cm}}{r}\right) \!\!\left(\frac{1\,\text{GHz}}{\nu_0}\right).
\end{equation}

Even for modest burst luminosities $L \sim 10^{38}$–$10^{40}\,\text{erg s}^{-1}$ at distances $r \sim 10^{10}$–$10^{11}\,\text{cm}$ where the unmagnetized approximation is better justified, we find $a_0 \sim 10$–$10^3$, well above unity. Meanwhile, the single-species plasma frequency in the wind satisfies $\omega_p/\omega_0 \sim 10^{-2}$–$10^{-1}$ for typical pair multiplicities $\kappa \sim 10^3$–$10^5$ and surface fields $B_s \sim 10^{14}\,\text{G}$ \cite{Sobacchi2022, Beloborodov2021}, placing FRBs in the underdense regime. FRBs therefore realize the strongly nonlinear parameter space $a_0 \gg 1$, $\omega_p < \omega_0$ that is the focus of this work.
Several physically important consequences follow from the large $a_0$ analysis presented here. 
First, the growth rate of the stimulated backscatter instability, 
shown in Eqns. (\ref{eq:growth-rate-LP}) and (\ref{eq:growth-rate-CP}) for LP and CP peak near $a_0$ of 2. At larger $a_0$ the growth rate decreases as $a_0^{-2/3}$ (Eq. 50) owing to the increasing relativistic mass of the drifting pairs and the Doppler shift due to the drift. Nevertheless, even at $a_0 \sim 10^2$–$10^3$ the implied e-folding timescale $({\lesssim}\,\mu\text{s})$ remains orders of magnitude shorter than a typical FRB pulse duration of $\sim 1\,\text{ms}$, so the instability has ample time to grow to nonlinear amplitudes within a single burst. Second, it is shown that the parameter $\epsilon_p$ is indeed important for large values of $a_0$. When $\epsilon_p>1$ the incoming wave will etch away by tranferring nearly $100\%$ of its energy to accelerating the pair plasma forward. Thus, in this case there is no reflection. 
Third, as discussed in the Appendix, the pump wave significantly depletes the ambient pair proper density. This occurs for both LP and CP but it is easier to analyze for CP. It was shown that 
when $n^{(0)}/n_0 \sim (a_0\omega_p/\omega_0)^{-2/3} \ll 1$  the effective plasma frequency of the depleted equilibrium 
can be far below $\omega_p(n_0)$. Equivalently, the relativistic critical density $n_{c,\text{rel}} = a_0 n_c$ can exceed the ambient density by orders of magnitude, allowing the FRB and its backscattered companion to propagate freely through regions that are classically overdense. 
Last, for linearly polarized FRBs, as opposed to circularly polarized, the parametric coupling excites all odd harmonics of the fundamental Doppler-shifted frequency. These produce a spectral comb whose spacing contracts as $1/a_0^2$ with increasing pump amplitude, potentially observable as sub-pulse frequency structure at high time resolution.
These estimates indicate that FEL-type stimulated backscatter is a physically plausible and potentially significant process during FRB propagation through magnetar winds, extending the earlier weak-pump analyses of Refs \cite{Ghosh2022, Iwamoto2026} into the strongly nonlinear regime directly relevant to observed FRB intensities. A fully quantitative prediction requires coupling the present theory to a realistic wind model accounting for finite pair temperature, magnetization, and the three-dimensional geometry of the propagation path, which we leave to future work.

\section{Summary}
A relativistic theory of intense EM propagation in underdense pair plasma was derived and compared to PIC simulation results. 
Utilizing the formalism from \cite{Sprangle-FEL} and insight from \cite{XuMori2026, decker1996evolution},  
we derived the dispersion for a general elliptically polarized EM pulse impinging on a pair plasma. It was shown that a linearly polarized EM pulse is susceptible to stimulated backscatter at the odd harmonics of the Doppler shifted laser frequency. This theory requires deriving the equilibrium for the pair plasma under the influence of an arbitrarily large field. The assumptions of the analysis were carefully dilineated and the theory was validated against OSIRIS PIC simulations. The simulations confirmed the instability growth rate and showed that multiple odd harmonics can be generated.   
The effects of finite temperature was  studied  via PIC simulations, from which it was found that the growth of backscatter gradually decreases as the temperature approaches keV and that harmonics $l > 0$ gradually disappear.  

\section{Acknowledgements}
This work was supported under the U.S. Department of Energy through  LLE subcontracts SUB000001031 and SUB00000211. The work of T.G. was supported FCT (Portugal) through  Grant No. CEECIND/04050/2021.
The work of J.P.P. was supported by the U.S. Department of Energy National Nuclear Security Administration under Award No. DE-NA0004144. 

\section{Appendix}
\subsection{Finite density effects}
\begin{figure}[H]
     \centering
     \begin{subfigure}[b]{0.4\textwidth}
         \centering
         \includegraphics[width=\textwidth]{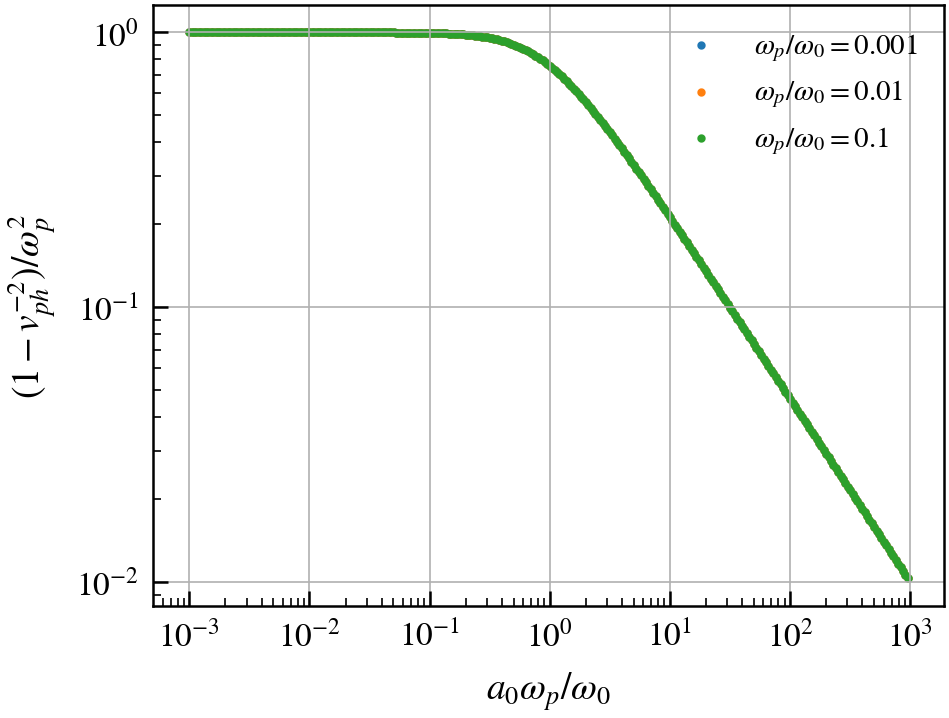}
     \end{subfigure}
     \caption{Pump dispersion relation vs $a_0\omega_p/\omega_0$}
     \label{fig:fin-dens-vph}
\end{figure}

\begin{figure}[H]
     \centering
     \begin{subfigure}[b]{0.4\textwidth}
         \centering
         \includegraphics[width=\textwidth]{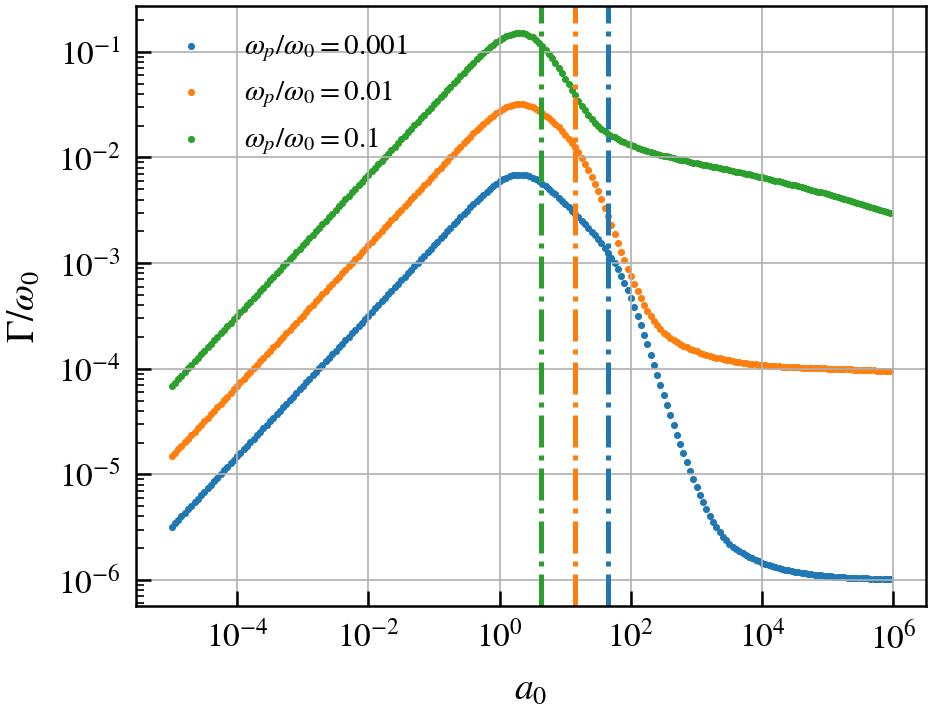}
     \end{subfigure}
     \begin{subfigure}[b]{0.4\textwidth}
         \centering
         \includegraphics[width=\textwidth]{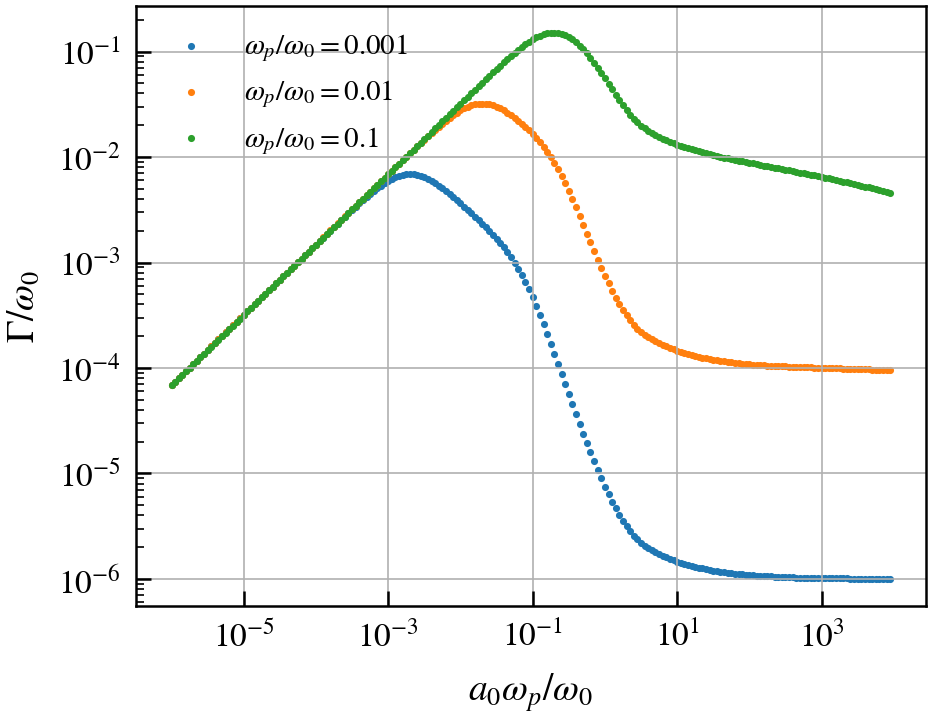}
     \end{subfigure}
     \caption{Maximal growth rate vs $a_0$ (left) and vs $a_0\omega_p/\omega_0$ (right) for circularly polarized pump at various $\omega_p/\omega_0$. The vertical dashed lines on the left are drawn at the $a_0$ where the lab frame frequency of the backscattered light wave in the underdense approximation becomes less than $\omega_p$.}
     \label{fig:fin-dens-disp}
\end{figure}

Although a complete perturbative analysis for any $a_0 \omega_p / \omega_0$ and polarization is beyond the scope of this work, we can still extract useful information on the behavior of the instability in the finite density case where the pump and/or the reflected light do not have phase velocities equal to c if we assume that the pump wave is circularly polarized. We expect this physics to become important at large enough $a_0$ such that the frequency of the backscattered light becomes smaller than $\omega_p$ so that it is not obvious that  
it can propagate in the plasma. In the circularly-polarized case the harmonic generation is absent, and the equilibrium solutions are constant, which simplifies the analysis considerably. The zeroth-order solutions are got by assuming the fluid variables are functions of $\xi = x - v_{ph} t$ where $v_{ph} = \omega_0 / k_0$. In terms of this variable, the fluid equations \ref{eq:9} are modified to take the form
\begin{align}
    \p_\xi(-v_{ph}\gamma + u_x )n &= 0\\
    (v_{ph}\gamma-u_x)\p_\xi u_x &= \frac{1}{2}\p_\xi\mathbf{a}^2\\
     (v_{ph}\gamma-u_x)\p_\xi \gamma &= \frac{v_{ph}}{2}\p_\xi\mathbf{a}^2\\
     (v_{ph}^2-1)\p_\xi^2 \mathbf{a} &= -\omega_p^2\frac{n}{n_0}\mathbf{a}\label{eq:finite-density-equil}
\end{align}
The corresponding solutions for $n,u_x,\gamma$ are:
\begin{align}
    n^{(0)} &= \frac{n_0}{\gamma^{(0)} - u_x^{(0)} /v_{ph}}\\
    \gamma^{(0)} &= 1 + v_{ph}u_x^{(0)}\\
    u_x^{(0)} &= -\frac{v_{ph} + \sqrt{v_{ph}^2 + a_0^2(v_{ph}^2 - 1)}}{v_{ph}^2 - 1}
\end{align}
These equations combined with the light wave equation yields the following dispersion relation for the (circularly-polarized) pump:
\begin{align}
    (1 - v_{ph}^{-2}) &= \frac{\omega_p^2}{\omega_0^2}\frac{n^{(0)} }{n_0}
\end{align}
A plot of the $v_{ph}$ vs $a_0$ is shown in \ref{fig:fin-dens-vph}
Even at various $\omega_p$, it is evident that the dispersion relation only depends on the parameter $a_0\omega_p/\omega_0$. A fact that is not obvious from the equilibrium solutions. Furthermore, that the effective plasma frequency decreases dramatically with increasing $a_0\omega_p / \omega_0$ means that the backscattered light wave can still propagate through the plasma even after the frequency shifts below critical.
The stability analysis proceeds by perturbing about the fluid variables as $n = n^{(0)} + \delta n, \gamma = \gamma^{(0)} + \delta \gamma, u_x = u_x^{(0)} + \delta u_x, \mathbf{a} = \mathbf{a}^{(0)} + \delta\mathbf{a}$. Extracting the first-order variation of the above fluid equations yields:
\begin{align}
    (\p_t + v_d\p_x)\frac{\delta n}{n_0} &= -\frac{1}{\gamma^{(0)}}\p_x\delta u_x - \frac{1}{\gamma^{(0)}}\p_t \delta \gamma\\
     (\p_t + v_d\p_x)\delta \gamma &= \frac{1}{\gamma^{(0)}}\p_t(\mathbf{a}^{(0)}\cdot \delta\mathbf{a}^*)\\
     (\p_t + v_d\p_x)\delta u_x &= -\frac{1}{\gamma^{(0)}}\p_x(\mathbf{a}^{(0)}\cdot \delta\mathbf{a}^*)\\
     (\p_t^2 -\p_x^2+\omega_p^2\frac{n^{(0)}}{n_0})\mathbf{\delta a} &= -\omega_p^2\frac{\delta n^*}{n_0}\mathbf{a}^{(0)}
\end{align}
where $v_d$ is the electron/positron drift velocity, $u_x^{(0)}/\gamma^{(0)}$. At this stage, it is convenient to assume that the backscattered light wave $\delta \mathbf{a}$, has the plane wave form of $e^{i(k_1x - \omega_1t} $ and is polarized in the same direction as the pump so that $\delta\mathbf{a}^*\cdot \mathbf{a}_0 = a_0 a_1 e^{i(kx - \omega t)}$ where $\omega,k \equiv \omega_0 - \omega_1,k_0 - k_1$. The first three equations can be readily inverted in Fourier space:
\begin{align}
    \delta \gamma &= \frac{a_0a_1}{\gamma^{(0)}}\frac{\omega}{-\omega + v_d k}\\
    \delta u_x &= \frac{a_0a_1}{\gamma^{(0)}}\frac{k}{-\omega + v_d k}\\
    \frac{\delta n}{n^{(0)}} &= \frac{a_0a_1}{(\gamma^{(0)})^2}\frac{\omega^2 - k^2}{(\omega - v_d k)^2}
\end{align}
Inserting the expression for $\delta n /n^{(0)}$ into the light wave equation yields the lab-frame dispersion relation for the stimulated backscatter:
\begin{align}
    -(\omega_0 - \omega)^2 +k_1^2 + \omega_p^2\frac{n^{(0)}}{n_0} &= \frac{\omega_p^2a_0^2}{2(\gamma^{(0)})^2}\frac{\omega^2 - k^2}{(\omega - v_d k)^2}\label{eq:disp-dens}
\end{align}
The extra factor of $1/2$ on the RHS comes from the normalization $|\delta{a}|^2= 2 a_1^2$. Note that $\gamma^{(0)}$ has an additional dependence on $a_0$. In the underdense limit, the imaginary part of $\omega$ is maximal when $k_1$ is equal to minus the incident wavenumber \textit{in the drift frame}. This is not true in the general case, however, and the maximum growth rate must be computed by numerically solving \ref{eq:disp-dens} for $\omega,k$. Plots of the full growth rate as functions of $a_0$ and $\omega_pa_0/\omega_0$ are shown in Fig.\ref{fig:fin-dens-disp}. The dashed lines in the left plot show where the backscattered frequency dips below $\omega_p$. This provides a rough measure of where the underdense limit no longer holds.
\bibliography{references} 


\end{document}